\documentclass[journal]{IEEEtran}
\IEEEoverridecommandlockouts
\usepackage{cite}
\usepackage{amsmath,amssymb,amsfonts}
\usepackage{mathtools}
\usepackage[ruled,vlined]{algorithm2e}
\usepackage{graphicx}
\usepackage{textcomp}
\usepackage[table]{xcolor}
\usepackage{hyperref}
\usepackage[nolist]{acronym}
\usepackage{tabularx}
\usepackage{booktabs}
\usepackage{soul}
\usepackage{makecell}
\usepackage{etoolbox}
\usepackage{titlesec}

\titlespacing{\subsection}{3pt}{*0.05}{*0.1}
\titlespacing{\subsubsection}{\parindent}{*0.05}{*0.1}
\begin{acronym}
    \acro{AWGN}{additive white Gaussian noise}
    \acro{ISAC}{integrated sensing and communication}
    \acro{DFRC}{dual-function radar communication}
    \acro{AoA}{angle of arrival}
    \acro{PR}{passive radio}
    \acrodefplural{AoA}[AoAs]{angles of arrival}
    \acro{ToA}{time of arrival}
    \acrodefplural{ToA}[ToAs]{times of arrival}
    \acro{AoD}{angle of departure}
    \acrodefplural{AoD}[AoDs]{angles of departure}
	\acro{LoS}{line-of-sight}
	\acro{NLoS}{non-line-of-sight}
	\acro{ULA}{uniform linear array}
	\acro{OFDM}{orthogonal frequency-division multiplexing}
	\acro{CRB}{Cram\'er-Rao bound}
	\acro{i.i.d}{independently and identically distributed}
	\acro{DFT}{discrete Fourier transform}
	\acro{SISO}{single-input, single-output}
    \acro{MIMO}{multiple-input, multiple-output}
	\acro{MSE}{mean squared error}
    \acro{CM}{channel modeling}
	\acro{SNR}{signal-to-noise ratio}
    \acro{SINR}{signal-to-interference-and-noise ratio}
	\acro{BGG}{bistatic geometric gain}
	\acro{ESNR}{estimation signal-to-noise ratio}
	\acro{DMC}{dense multipath components}
	\acro{SC}{specular components}
	\acro{RCS}{radar cross section}
	\acro{TF}{transfer function}
	\acro{PDP}{power-delay profile}
	\acro{PAP}{power-angular profile}
	\acro{WB}{wideband}
	\acro{UWB}{ultra-wideband}
	\acro{SV}{Saleh-Valenzuela}
	\acro{AS}{angular spread}
	\acro{DS}{delay spread}
	\acro{FCF}{frequency correlation function}
	\acro{VMD}{Von-Mises distribution}
	\acro{FIM}{Fisher information matrix}
	\acro{LM}{Levenberg-Marquardt}
	\acro{MRC}{maximal ratio combining}
	\acro{RMSE}{root mean squared error}
	\acro{Tx}{transmitter}
	\acro{Rx}{receiver}
	\acro{THz}{terahertz}
	\acro{UE}{user equipment}
	\acro{BS}{base station}
	\acro{PD}{probability of detection}
	\acro{PFA}{probability of false alarm}
	\acro{CSI}{channel state information}
	\acro{FFT}{fast Fourier transform}
	\acro{MCRB}{misspecified \ac{CRB}}
	\acro{JADE}{joint angle and delay estimation}
\end{acronym}

\definecolor{gray10pct}{gray}{0.9}

\DeclareMathOperator{\toep}{toep}

\DeclareMathOperator{\CRB}{CRB}
\DeclareMathOperator{\MCRB}{MCRB}

\DeclareMathOperator{\cat}{cat}
\DeclareMathOperator{\NF}{NF}

\DeclareMathOperator{\var}{var}
\DeclareMathOperator{\PD}{PD}
\DeclareMathOperator{\PFA}{PFA}
\DeclareMathOperator{\BW}{BW}

\DeclareMathOperator*{\argmin}{arg\,min}
\DeclareMathOperator*{\blockdiag}{blkdiag}

\newcommand{\norm}[1]{\left\lVert#1\right\rVert}
\newcommand{\magn}[1]{\left\lvert#1\right\rvert}

\newif\iftrackedchanges
\trackedchangesfalse   

\usepackage{xcolor}
\DeclareRobustCommand{\oscar}[1]{{%
  \iftrackedchanges\color{blue}\fi
  #1%
}}

\makeatletter

\newcommand{\revisionreference}[1]{%
  \expandafter\def\csname revisionreference@#1\endcsname{}%
}

\newcommand{\revisionreferencecolor}[1]{%
  \color{black}%
  \iftrackedchanges
  	\ifcsname revisionreference@#1\endcsname
      \color{blue}%
  	\fi
  \fi
}

\revisionreference{Pegoraro2026MultibandFR3}
\revisionreference{Pegoraro2024HiSAC}
\revisionreference{RichmondHorowitz2015}
\revisionreference{KhatibEtAl2026}

\pretocmd{\@bibitem}{\revisionreferencecolor{#1}}{}{}
\pretocmd{\@lbibitem}{\revisionreferencecolor{#2}}{}{}

\makeatother

\def\BibTeX{{\rm B\kern-.05em{\sc i\kern-.025em b}\kern-.08em
    T\kern-.1667em\lower.7ex\hbox{E}\kern-.125emX}}
\begin{document}

\bstctlcite{IEEEexample:BSTcontrol}

\title{Multi-Band Sensing in FR3 with Background Dense Multipath Components}

\author{Dexin Wang, Roberto Bomfin, Ahmad Bazzi, and Marwa Chafii
\thanks{This work has been supported in part by Tamkeen under the Research Institute NYUAD Grant CG017, and in part by the New York University Abu Dhabi (NYUAD) Center for Artificial Intelligence and Robotics, funded through Tamkeen under the Research Institute NYUAD Award CG010.}
\thanks{Dexin Wang, Ahmad Bazzi, and Marwa Chafii are with the Engineering Division, NYUAD, P.O. Box 129188, Abu Dhabi, UAE and NYU WIRELESS, New York University (NYU) Tandon School of Engineering, Brooklyn, NY 11201, USA (email: \mbox{oscarwang@nyu.edu}, \mbox{ahmad.bazzi@nyu.edu}, \mbox{marwa.chafii@nyu.edu}).}
\thanks{Roberto Bomfin is with the Engineering Division, NYUAD, P.O. Box 129188, Abu Dhabi, UAE (email: \mbox{roberto.bomfin@nyu.edu}).}
}

\markboth{Accepted for publication in IEEE Transactions on Wireless Communications, 2026}%
         {Accepted for publication in IEEE Transactions on Wireless Communications, 2026}
\maketitle

\begin{abstract}
Multi-band sensing has emerged as a key enabler of integrated sensing and communication (ISAC), one of the six primary usage scenarios defined for IMT-2030 (6G). The introduction of frequency range 3 (FR3, 7-24 GHz), comprising non-contiguous sub-bands across a wide frequency span, further reinforces the importance of multi-band operation. In such scenarios, frequency-dependent propagation effects that are collectively referred to as dense multipath components (DMC), including clutter, diffraction, and diffuse scattering, must be carefully considered. Building on prior literature and our experimental observations, this paper proposes a novel ISAC channel analysis tailored to multi-band sensing, based on a channel model with background DMCs. It also assesses the sensing trade-offs among sub-bands by analyzing Cram\'er-Rao bound (CRB)-based fundamental limits.
Furthermore, a scalable multi-band estimator is proposed that resolves angular ambiguities arising from the grating lobes effect. Simulation results of the multi-band estimator demonstrate substantial gains in estimation accuracy and reductions in false alarm rate over single-band estimators operating on each constituent sub-band within the CRB-achieving regime. In a representative test case, the proposed estimator achieves reductions of \oscar{37.4}\% and \oscar{17.2}\% in the root mean squared error of delay estimation compared to single-band estimators operating at 8.75 GHz and 21.7 GHz, respectively.
\end{abstract}


\section{Introduction}
\label{sec:introduction}
\IEEEPARstart{F}{\lowercase{uture}} 6G systems are envisioned to reliably facilitate novel applications that interact more deeply with the physical world, such as the Internet of Things, autonomous driving, extended reality, smart cities, digital twins \cite{11215841}, and radio maps for environment-aware communication \cite{11083758}. Thus, to save resources in bandwidth, hardware, processing power, and energy while facilitating efficient operation of such a large number of devices, \ac{ISAC} has been proposed as one of the six key usage scenarios of IMT-2030 (6G)\cite{chafii2023twelve,liu2025itu, liu2022integrated, gonzalez2024integrated}. One type of \ac{ISAC} is \textit{communication-centric \ac{ISAC}} \cite{11161464}, which investigates the problem of performing sensing via existing communication protocols and infrastructure \cite{10464930,11272093}, including the allocated frequency bands. With the addition of frequency range 3 (FR3, 7-24 GHz) \cite{bazzi2025uppermidbandspectrum6g, Rappaport2024InH_PointData, shakya2024meas} and \ac{THz} \cite{11153494}, future 6G systems will likely operate in a wide range of frequencies and even multiple sub-bands \cite{Mezzavilla2024frequencyhopping}. Hence, one interesting topic in 6G \ac{ISAC} will undoubtedly be to perform multi-band sensing in FR3 using signals from communication protocols \cite{raviv2024multi,wang2025compressed}.

\subsection{Background Concepts and Related Works}

\subsubsection{\oscar{Multi-Band Sensing}}
Multi-band sensing is the technique of performing sensing across non-contiguous frequency bands. From the literature, some of the reasons for choosing multi-band sensing are as follows \cite{wan2024ofdm,li2025multinode,hu2023mbrhs,hu2023twostagemb}. First, to achieve higher time resolution, a typical approach is to use a large bandwidth. Nevertheless, increasing bandwidth is often associated with higher hardware costs and greater algorithmic complexity. Second, the increased amount of sensing data implies a lower error probability. For example, if severe fading happens at a particular frequency band, information from other bands can still be used. For single-band \ac{ISAC} systems, where ideal wideband sensing waveforms, such as radar chirps, are typically unavailable, the loss can be mitigated by using multi-band sensing. Third, by accounting for frequency dependence in the signal model, we can estimate more environmental parameters and obtain additional information on the target, such as size, material, or roughness, beyond Cartesian locations alone. Finally, since the frequency bands allocated to one operator are typically non-contiguous (especially in FR3, considering coexistence with other occupiers such as satellites \cite{testolina2024sharing}), multi-band sensing is a practical assumption considering \ac{ISAC} applications. 

\oscar{Despite the many advantages outlined for employing multi-band sensing, we should note that the frequency anisotropy of targets, which was commonly not modeled in multi-band \ac{ISAC} works \cite{Pegoraro2026MultibandFR3}, can no longer be ignored in the high fractional bandwidth case of multi-band sensing in FR3. Frequency anisotropy poses several challenges. First, it invalidates existing \textit{coherent} multi-band algorithms and phase offset compensation techniques across bands \cite{Pegoraro2026MultibandFR3}. Second, grating lobe artifacts \textit{in the range domain} will be induced when multiple bands are aggregated \cite{Pegoraro2026MultibandFR3}.}

\subsubsection{\oscar{Dense Multipath Components (DMC)}}
\label{subsubsec:DMC}
Typical sensing processing in \ac{ISAC} literature relies on parameter estimation from \ac{SC}-only channel models, which are defined by the geometrical structures of the sensing targets in the environment, including \ac{ToA} delay, \ac{AoD}, \ac{AoA}, and Doppler, among others \cite{Fortunati_Massive_MIMO_Radar_2020, Chiriyath_Radar_Comm_Coexistence_2016, Kumari_80211ad_Radar_2018, Bomfin_Chafii_UW_ISAC_2024}. However, an important propagation effect widely studied in \ac{CM} literature, while remaining largely underexplored in \ac{ISAC}, is the \ac{DMC} \cite{molish2022DMCSurvey,  Richter_DMC_Tracking_2006, Kaske_Thoma_ML_DMC_2015, Hanssens_PAP_DMC_2018, Jost_Path_Detection_2012, Saito_DMC_Indoor_2016}. \oscar{\ac{DMC} refers to the unresolved multipath contribution formed by many weak components, such as diffuse scattering and diffraction  \cite{molish2022DMCSurvey}. Any physical object may produce both a resolvable coherent component and DMC.} For example, \ac{DMC} could rise from surface roughness and irregularities, building structure, or walls, etc. These propagation effects are frequency-dependent and should be modeled differently across sub-bands. Moreover, since \ac{DMC} are modeled as non-resolvable multipath components, their magnitude scale with transmit power and should hence be treated as interference.

\Ac{DMC} is a dominant impairment in practical multi-band \ac{ISAC} channels and must be explicitly captured in the system model to obtain meaningful sensing conclusions. In contrast to the common intuition, the reflected signal power alone is not a reliable proxy for parameter estimation performance once \ac{DMC} is present: \ac{DMC}, as interference, grows with the \ac{SC}, and may even shadow the weak \ac{SC} peaks, degrading sensing performance even at high received power. Neglecting \ac{DMC} leads to systematically misleading results: from a system-level perspective, it yields over-optimistic, unrealistic predictions of sensing performance. From a signal processing perspective, it introduces non-negligible estimation bias and unexpectedly large errors, and yields erroneous internal reliability metrics for the estimated parameters.

In multi-band \oscar{FR3} settings, \ac{DMC} becomes even more determinant \oscar{not only} because its statistics are frequency-dependent, \oscar{but also because FR3 lies in the middle range where bandwidths can be wide enough to resolve these dense components while the channel still remains rich scattering}. In our previous works \cite{bomfin2024experiment,bomfin2026icc}, we investigated the multi-band \ac{ISAC} channel through single-band processing (i.e., without cross-band combination) and experimentally observed a \oscar{key} behavior: lower bands, although often enjoying higher received power due to less blockage, can suffer from stronger \ac{DMC} interference. So, the apparent power advantage does not necessarily translate into better sensing performance. Conversely, higher bands can exhibit sparser channels but with relatively weaker \ac{DMC} contamination, so that they may achieve superior estimation due to a higher \ac{SC}/\ac{DMC} power ratio even at lower \ac{SNR}. These effects are inherently non-linear and can violate conventional single-band intuition, making a proper \ac{DMC}-aware model essential for assessing the contributions of different sub-bands in multi-band \ac{ISAC} performance, \oscar{and for exploiting the relative coverage and reliability advantages of lower and higher sub-bands, respectively}.

\subsection{This Work}

Existing works typically fail to bridge the gap between multi-band sensing and \ac{DMC}. For example, multi-band sensing works such as \cite{wan2024ofdm,li2025multinode,hu2023mbrhs,hu2023twostagemb} all assumed an \ac{SC}-only channel model with white Gaussian noise, while \ac{DMC}-related works such as \cite{molish2022DMCSurvey,  Richter_DMC_Tracking_2006, Kaske_Thoma_ML_DMC_2015, Hanssens_PAP_DMC_2018, Jost_Path_Detection_2012, Saito_DMC_Indoor_2016} are focused on \ac{CM} and single-band estimation of channel parameters. Moreover, despite the \ac{DMC} and multi-band sensing-related insights provided by our previous works \cite{bomfin2024experiment,bomfin2026icc}, they were limited to single-band estimation of  \ac{SC} under \ac{DMC}. The theoretical framework for \textit{combined} multi-band processing with sub-band frequency-dependent \ac{DMC} remains unaddressed. Additionally, accounting for the justification of including \ac{DMC} in multi-band \ac{ISAC} channel models discussed in the previous subsection, all these reasons motivate the present work: to analyze the effects of \ac{DMC} on combined multi-band \ac{ISAC} systems and compare its estimation and detection performance with single-band processing.

Our contributions in this paper are summarized as follows. First, to understand the impact of \ac{DMC} in multi-band sensing:
\begin{itemize}
\item We propose and analyze a novel \ac{ISAC} system model for multi-band sensing involving background \ac{DMC}.
\item We show how the nonlinearities brought by \ac{DMC} challenge conventional sensing intuitions through plots of the \ac{CRB} and the modified Cassini ovals obtained from our proposed \ac{BGG} metric.
\item We assess the sensing tradeoffs among sub-bands by analyzing \ac{CRB}-based fundamental limits, showing how one sub-band may no longer dominate sensing performance under all conditions.
\end{itemize}
Then, to know how to perform multi-band sensing under \ac{DMC}:
\begin{itemize}
\item We propose a multi-band estimator capable of resolving angular ambiguity caused by the grating lobes effect.
\item We observe fundamental limits-achieving performances, with enhanced performance compared to single-band systems in terms of lower \ac{RMSE} and \ac{PFA}.
\end{itemize}

The remainder of this paper is organized as follows. Section \ref{sec:system-model} describes the multi-band system model, incorporating both the \ac{SC} and \ac{DMC}. Section \ref{sec:fundamental-limits-and-empirical-metrics} defines the fundamental limits and empirical metrics that help understand the non-linear impact of \ac{DMC} on the estimation accuracy and detectability of the paths. Section \ref{sec:proposed-multi-band-estimation-algorithm} details the design of our scalable multi-band estimator that resolves angular ambiguity from the grating lobes effect. Section \ref{sec:numerical-results} reports and analyzes the numerical results for the system model and the algorithm. Finally, Section \ref{sec:conclusions-and-future-work} concludes the paper and outlines directions for future research.

\textbf{Notations}: $\otimes$ denotes the Kronecker product, $\odot$ denotes the element-wise product, and $\toep(\pmb{x})$ denotes the Toeplitz matrix where the first column is $\pmb{x}$ and the first row is $\pmb{x}^H$. 
Moreover, $\cat_n\{\cdot\}$ denotes concatenating tensors in the set $\{ \cdot \}$ in the $n^{\text{th}}$ dimension.
Also, $\{ a_n \}_{n=1}^{N}$, which may also be written as $a_{1:N}$, denotes the set $\{ a_1,a_2,\dots,a_N \}$, and $[ \pmb{a}_n ]_{n = 1}^N$ denotes the vertically stacked vector $[\pmb{a}_1^T,\pmb{a}_2^T,\dots,\pmb{a}_N^T ]^T$.
Additionally, $\pmb{0}_{\times n}$ denotes a matrix of all zeros with $n$ columns if the number of rows can be inferred from other parts of the equation.
Finally, for some matrix $\pmb{A}$, $\left[\pmb{A}\right]_{i,:}$ denotes the $i^{\text{th}}$ row, $\left[\pmb{A}\right]_{:,j}$ denotes the $j^{\text{th}}$ column, and $\left[\pmb{A}\right]_{i,j}$ denotes the $(i,j)^{\text{th}}$ element.

\section{System Model}
\label{sec:system-model}
\begin{figure}[!t]
\centering
\includegraphics[width=3.5in]{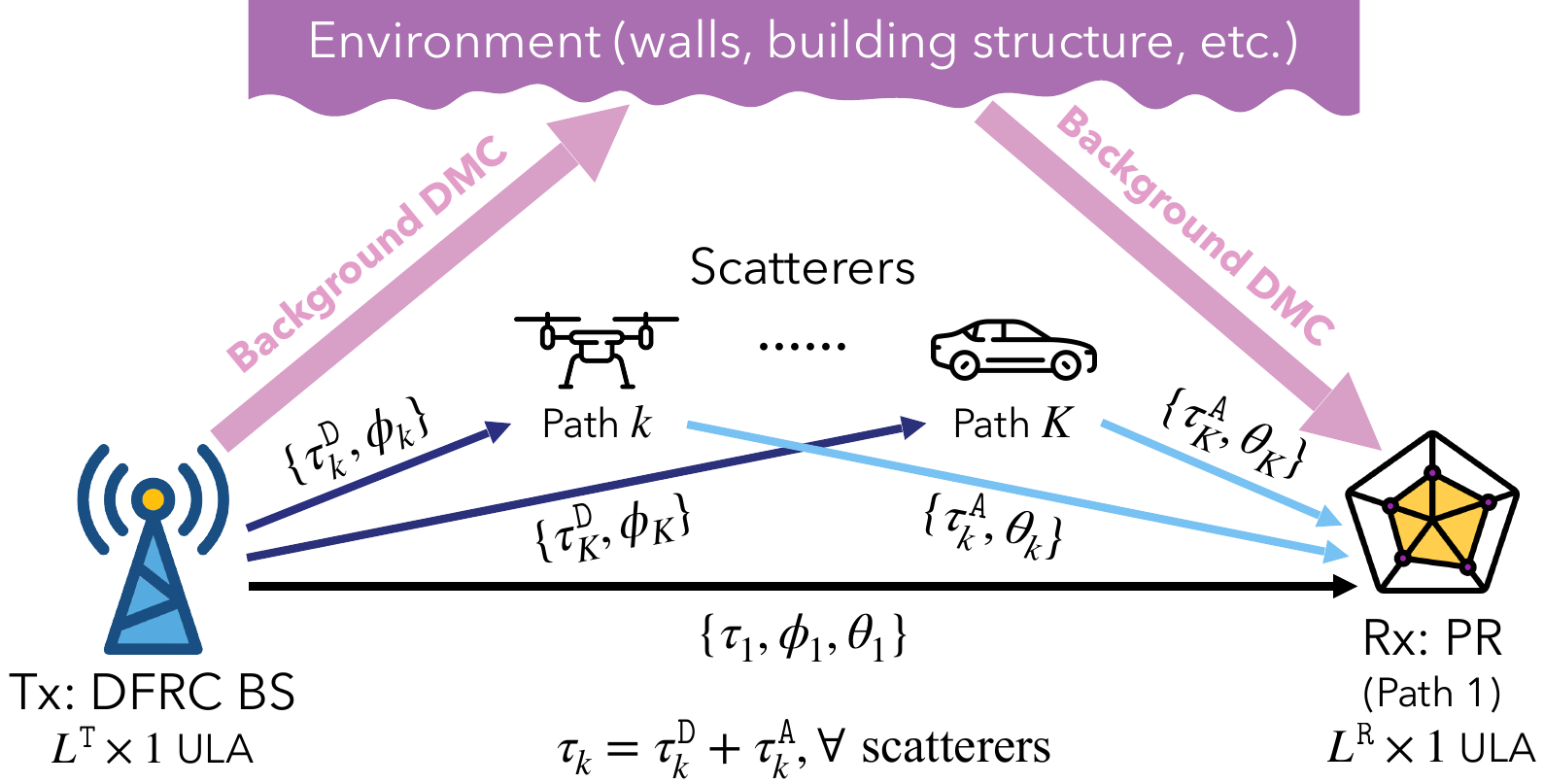}
\caption{Bi-static ISAC scenario with $K-1$ scatterers under background \ac{DMC} interference. The geometric sensing parameters associated with each path are shown.}
\label{fig:scenario}
\end{figure}

The scenario, as shown in Fig. \ref{fig:scenario}, involves a \ac{DFRC} \ac{BS} serving as the \ac{Tx} and a \ac{PR} as the \ac{Rx}, whose Cartesian locations are assumed to be known. The \ac{Tx} \ac{DFRC} \ac{BS} has $L^{\tt{T}}$ \ac{ULA} antenna elements, and the \ac{Rx} \ac{PR} has $L^{\tt{R}}$ \ac{ULA} antenna elements.  The \ac{PR} receives the multi-band signal from the \ac{BS} across $M$ sub-bands, each with $N$ subcarriers, and performs bi-static sensing using the downlink preamble signals from communications with \acp{UE} in the environment. Note that in this paper, all sub-bands are assumed to be identical in terms of the number of subcarriers, even though this may not be the case in general. The channel consists of $K-1$ scatterers and the \ac{DMC}. 
In addition to the scatterers, the \ac{LoS} path is also estimated. In general, the \ac{LoS} path serves as a reference for localizing scatterer positions when the \ac{Tx} and \ac{Rx} are not time-synchronized, and their respective orientations are unknown.
So, accounting for both the \ac{LoS} and the scatterer paths, the total number of paths is $K$. \oscar{We assume each of the $K$ paths to be \ac{SC}-only, meaning that they are resolvable and coherent. The present model focuses on aggregating all \ac{DMC} as a background nuisance component, so target-dependent diffuse returns are not separately parameterized.}

In this paper, we study the typical \ac{ISAC} system in which sensing is performed on the \ac{MIMO} frequency-domain \ac{CSI}. In practice, the \ac{CSI} is obtained via \ac{OFDM} pilots that are designed to ensure orthogonality across \ac{Tx} antennas. Following the model in \cite{molish2022DMCSurvey}, we express the multi-band \ac{MIMO} channel \ac{TF} $\pmb{h} \in \mathbb{C}^{NL^{\tt{T}}L^{\tt{R}}M \times 1}$ by
\begin{equation}
\label{eq:channel_tf}
	\pmb{h} = \pmb{s} + \pmb{d} + \pmb{w},
\end{equation}
where $\pmb{s} = \left[{\pmb{s}_{m}}\right]_{m=1}^M$ represents the \ac{SC} associated with the $K$ paths, $\pmb{d}\sim\mathcal{CN}\left(\pmb{0},\pmb{R}\right)$ represents the background \ac{DMC}, and $\pmb{w}\sim\mathcal{CN}\left(\pmb{0},\pmb{S}\right)$ represents the noise with covariance matrix $\pmb{S}$. We use $m$ to denote the sub-band index, \oscar{and we use the term \ac{SC} in its conventional \ac{CM} sense, which refers to any discrete and resolvable multipath components that are aggregated from phase-coherent contributions, rather than exclusively mirror-like geometric-optics reflections. Thus, scattering, diffraction, or reflection may contribute to an \ac{SC} when they produce a resolvable coherent path, and contributions that aggregate non-coherently and cannot be individually resolved are instead included in the \ac{DMC} term. Consequently, the same wall, vehicle, or any other object may physically contribute to both terms.} We expect the \ac{RCS} of the scatterers, the antenna radiation patterns, and the background \ac{DMC} properties to be frequency-dependent across sub-bands, while remaining constant within each sub-band. The detailed modeling approaches for each of the three parts (\ac{SC}, \ac{DMC}, and noise) are discussed in the following subsections.

\subsection{The Specular Components (SC)}

For a fixed environment (i.e., placement of Tx, Rx, and the scatterers) over the entire signal duration, the \ac{SC} is generally considered to be time-invariant, i.e., not changing across different realizations of the channel. Hence, each $\pmb{s}_{m} \in \mathbb{C}^{NL^{\tt{T}}L^{\tt{R}} \times 1}$ represents the specular channel \ac{TF} for each sub-band with frequency $f_m$, which is given by
\begin{equation}
\label{eq:h_per_subcarrier}
\pmb{s}_{m} = 
\sum_{k=1}^K g_{k,m} \pmb{a}_{m}(\tau_k,\phi_k,\theta_k),
\end{equation}
where $\pmb{a}_{m}(\tau_k,\phi_k,\theta_k)$ is the total steering vector for path $k$ at sub-band $m$ described by the geometric parameters, and $g_{k,m}$ is the path gain for path $k$ at sub-band $m$.

\subsubsection{Geometric Parameters}
The geometric parameters of the $k^{\textrm{th}}$ path include the \ac{ToA} delay $\tau_k$ relative to the \ac{LoS}, the \ac{AoD} $\phi_k$ relative to the \ac{Tx} array broadside, and the \ac{AoA} $\theta_k$ relative to the \ac{Rx} array broadside. They contribute to the \ac{SC} part of the channel model through the total steering vector, which can be factorized as \cite{molish2022DMCSurvey}
\begin{equation}
\pmb{a}_{m}(\tau_k,\phi_k,\theta_k) = 
\pmb{a}^{\tt{F}}_m(\tau_k) \otimes 
\pmb{a}^{\tt{T}}_m(\phi_k) \otimes 
\pmb{a}^{\tt{R}}_m(\theta_k).
\end{equation}
The path index $k$ is designated as follows: $k=1$ corresponds to the \ac{LoS} path, and $k=2 \cdots K$ corresponds to the scatterers. 
Assuming typical far-field propagation and narrowband antenna array response \cite{molish2022DMCSurvey}, the factorized terms of the \ac{SC} steering vector can be expressed as
\begin{equation}
\label{aF}
	\pmb{a}^{\tt{F}}_m(\tau) = 
	\left[
		\exp\left(
			-jn\omega_{\Delta,m} \tau
		\right)
	\right]_{n=0}^{N-1},
\end{equation}
\begin{equation}
\label{aT}
	\pmb{a}^{\tt{T}}_m(\phi) =
	\left[
		\exp\left(
			{-j \omega_m (d^{\tt{T}}/c)}
			\ell
			\sin\phi
		\right)
	\right]_{\ell = 0}^{L^{\tt{T}}-1},
\end{equation}
\begin{equation}
\label{aR}
	\pmb{a}^{\tt{R}}_m(\theta) =
	\left[
		\exp\left(
			{-j \omega_m (d^{\tt{R}}/c)}
			\ell
			\sin\theta
		\right)
	\right]_{\ell = 0}^{L^{\tt{R}}-1},
\end{equation}
where $\omega_{m} = 2\pi f_m$, $\omega_{\Delta,m} = 2\pi f_{\Delta,m}$ is the subcarrier spacing, and $d^{\tt{T}}$ and $d^{\tt{R}}$ are the antenna spacings of the \ac{Tx} and \ac{Rx} \ac{ULA}, respectively. 

\subsubsection{Path Gains}
In multi-band sensing, ideally, the same hardware is used across different sub-bands. Thus, we assume fixed-aperture antennas instead of fixed-gain ones, leading to varying antenna directivity patterns across sub-bands. Additionally, the \acp{RCS} of the scatterers are highly frequency-dependent. \oscar{Regarding the proper treatment of this frequency anisotropy, we model it only across different sub-bands given that we use narrow sub-bands.} Accurately modeling every single detail of the path gains requires a high degree of complexity due to the large number of variables, which compromises generalizability and is not the focus of this paper. Thus, we opted for the following model, which absorbs many parameters, including $\lambda_m^2/4\pi$, the \ac{RCS} of the scatterers $\sigma^{\tt{RCS}}_{k,m} (\forall k>1)$, the \ac{Tx} and \ac{Rx} antenna directivity patterns $u^{\tt{T}}_m(\phi)$ and $u^{\tt{R}}_m(\theta)$, and phase terms such as $e^{-j\omega_m\tau_k}$ into the complex path coefficient $\gamma_{k,m}$.

The gain of the \ac{LoS} path ($k=1$) is modeled by the Friis' equation
\begin{equation}
\label{gainLoS}
	g_{1,m} = \gamma_{1,m} 
	\sqrt{\frac{P^{\tt{T}}f_{\Delta,m}} 
	{4\pi(c\tau_1)^2}},
\end{equation}
where
\begin{equation}
	\magn{\gamma_{1,m}} = 
	\sqrt{\frac{\lambda_m^2 u^{\tt{T}}_m(\phi) u^{\tt{R}}_m(\theta)}{4\pi}},
\end{equation}
and the complex path gain for each scatterer's \ac{SC} $g_{k,m}\,(k>1)$ is modeled by the two-way bistatic radar equation
\begin{equation}
	g_{k,m} = \gamma_{k,m}
	\sqrt{\frac{P^{\tt{T}} f_{\Delta,m} }
	{(4\pi)^2 (c\tau_k^{\tt{D}})^2 (c\tau_k^{\tt{A}})^2 }},
\end{equation}
where
\begin{equation}
\label{xiscat}
	\magn{\gamma_{k,m}} = 
	\sqrt{\frac{\lambda_m^2 \sigma^{\tt{RCS}}_{k,m} u^{\tt{T}}_m(\phi) u^{\tt{R}}_m(\theta)}{4\pi}}.
\end{equation}
In the above equations \eqref{gainLoS}-\eqref{xiscat}, $P^{\tt{T}}$ is the fixed average transmit power spectral density across symbols, $\tau^{\tt{D}}_k$ is the delay between the \ac{Tx} and the scatterer, $\tau^{\tt{A}}_k$ is the delay between the scatterer and the \ac{Rx}, and $\gamma_{k,m}$ is the frequency-dependent complex path coefficient ($k=1$ for the \ac{LoS} path). The average transmit power per subcarrier across symbols on the $m^{\textrm{th}}$ subband is thus $P^{\tt{T}} f_{\Delta,m}$. \oscar{Note that these equations model the amplitude of each resolvable and coherent path's \ac{SC} through its frequency-dependent propagation and scattering response. Therefore, they do not require the underlying interaction to be a mirror-like reflection.}

For the purpose of this paper, it is also sufficient to capture the typical case where the path coefficient magnitude $\magn{\gamma_{k,m}}$ decreases with $f_m$. This assumption is consistent with two well-established physical trends. First, propagation losses associated with diffuse scattering, diffraction, and penetration through the scatterers generally increase at higher frequencies for most real-world scatterers with electrically rough surfaces, as reported in the \ac{UWB} and propagation physics literature \cite{molisch2009uwb, molisch2005uwb, haneda2012uwb, qiu1999multipath}. Second, for a fixed-size antenna, the electrical aperture grows with frequency, causing the radiation pattern to become more directive as $f_m$ increases \cite{balanis2016antenna}. According to antenna physics and the conservation of energy, when a path is not aligned with the boresight direction, the corresponding directivity typically oscillates with frequency but has a non-increasing envelope \cite{huang2021antennas}. This envelope trend contributes to a general frequency-dependent reduction in the effective gain observed by off-boresight paths.

\subsection{The Dense Multipath Components (DMC)}

Since \ac{DMC} arise from small surface details such as roughness, it may suffer from randomness across channel realizations due to small-scale mobility, i.e. translation or rotation of the Tx, Rx, and background scatterers on the order of wavelengths, which causes the phases of the individual \ac{DMC} in one resolvable delay/angle bin to vary randomly across an extended period of time even if the \ac{SC} stays constant. 
From the central limit theorem, in each delay bin, the \ac{DMC} is modeled as a complex Gaussian random variable due to the large number of unresolvable \ac{DMC} \cite{molish2022DMCSurvey}. A typical approach to modeling the background \ac{DMC} is to align it with the \ac{LoS} specular path \cite{molish2022DMCSurvey}. \ac{DMC} profiles supporting this modeling approach are observed and measured in our other experiment-based works \cite{bomfin2024experiment,bomfin2026icc}.

\subsubsection{Delay Domain Characteristics}
The \ac{PDP} of the complex Gaussian process can be modeled as an exponential decay \cite{rimaxthesis}:
\begin{equation}
P_m^{\tt{DS}}(\tau,\tau_1) =
\begin{cases} 
0 & \tau < \tau_1, \\
\alpha \magn{ g_{1,m} }^2 / 2 & \tau = \tau_1, \\
\alpha \magn{ g_{1,m} }^2 e^{-\beta_m(\tau - \tau_1)} & \tau > \tau_1,
\end{cases}
\end{equation}
where the maximum power of the \ac{DMC} is modeled as a fraction of the \ac{LoS} \ac{SC} power quantified by $\alpha$, and $\beta_m$ is the sub-band frequency-dependent decay rate. \ac{DMC} \acp{PDP} that can be described by this model are measured and observed in our experiment-based work \cite{bomfin2024experiment}.
By calculating the Fourier Transform of the \ac{PDP}, the \ac{FCF}, we can see that the decay rate can be interpreted as the coherence bandwidth of the \ac{DMC} on sub-band $m$, which contains the \ac{DS} information. The finite bandwidth sampled version of the \ac{FCF} is given by \cite{molish2022DMCSurvey}
\begin{equation}
\pmb{r}^{\tt{F}}_m = 
\frac{\alpha \magn{ g_{1,m} }^2}{N} 
\left[
 \frac{e^{-j n \omega_{\Delta,m} \tau_1}}{\tilde{\beta}_m + \frac{j 2\pi n}{N}}
\right]_{n=0}^{N-1},
\end{equation}
where $\tilde{\beta}_m = \beta_m/(N-1)\Delta\omega_m$ is the coherence bandwidth  normalized to the measurement bandwidth $(N-1)\Delta\omega_m$.
In the frequency domain, the complex Gaussian process is Fourier-transformed and sampled at the subcarriers, hence becoming joint complex Gaussian random variables following the distribution $\mathcal{CN}(\pmb{0},\pmb{R}_m^{\tt{F}})$, where $\pmb{R}_m^{\tt{F}} \in \mathbb{C}^{N\times N}$ is obtained from the \ac{FCF} and is given by 
\begin{equation}
\pmb{R}_m^{\tt{F}} = \toep(\pmb{r}^{\tt{F}}_m).
\end{equation}

\subsubsection{Angle Domain Characteristics}
 The \ac{PAP} can be modeled as a \ac{VMD} \cite{molish2022DMCSurvey}:
\begin{equation}
\label{PAP}
	P_m^{\tt{AS}}(\phi,\phi_1) = \frac{e^{\kappa_m\cos(\phi-\phi_1)}}{2\pi I_0(\kappa_m)},
\end{equation}
where $\kappa_m$ is the sub-band frequency-dependent \ac{VMD} parameter containing the \ac{AS} information.
Note that since we consider the maximum powers in the \ac{PDP}, we do not have to scale the \ac{PAP}.
To account for the contribution from each incremental \ac{AoD} and \ac{AoA}, an integration of the profile over the \ac{AS} is performed. 
We can write the angular covariance matrix of the \ac{DMC} as  
\begin{equation}
\label{RT}
	\pmb{R}^{\tt{T}}_m = \toep\left(\pmb{r}^{\tt{T}}_m\right),
\end{equation}
where 
\begin{equation}
\label{rT}
\pmb{r}^{\tt{T}}_m = \left[
\int_{-\pi}^{\pi}
\magn{u_m^{\tt{T}}(\phi)}^2
P_m^{\tt{AS}}(\phi,\phi_1)
\left[ \pmb{a}_m^{\tt{T}} \right]_{\ell}
d\phi \right]_{\ell = 1}^{L^{\tt{T}}}.
\end{equation}
Note that $\pmb{R}^{\tt{R}}_m$ follows a similar structure from \eqref{PAP} to \eqref{rT} by changing $\tt{T}$ to $\tt{R}$ and $\phi$ to $\theta$.

Now, based on the above definitions and assuming independence across different sub-bands, we can compactly write the \ac{DMC} covariance matrix as
\begin{equation}
\pmb{R} = \blockdiag\left\{\pmb{R}_m\right\}_{m=1}^M,
\end{equation}
where
\begin{equation}
\pmb{R}_m = 
\pmb{R}_m^{\tt{F}}
\otimes\pmb{R}_m^{\tt{T}}
\otimes\pmb{R}_m^{\tt{R}}
	\in
	\mathbb{C}^{NL^{\tt{T}}L^{\tt{R}} \times NL^{\tt{T}}L^{\tt{R}}}.
\end{equation}

Since \ac{DMC} is essentially multipath interference, they should suffer from higher diffuse scattering and diffraction as sub-band frequency increases, just like the \ac{SC}. In addition, we should also expect higher pathloss exponents for the \ac{DMC} as frequency increases, since the propagation environment for \ac{DMC} may not be free space. Hence, a practical way to model the \ac{DMC} is to assume higher decay rates $\tilde{\beta}$ and \ac{VMD} parameters $\kappa$ for higher frequency sub-bands. This trend of increasing decay rates with frequency is also experimentally verified in our other work \cite{bomfin2024experiment}.

\subsection{The Noise}
We assume the noise is white; thus, the noise covariance matrix is modeled as
\begin{equation}
	\pmb{S} = \blockdiag{\left\{\pmb{S}_m\right\}}_{m=1}^M
	= \blockdiag{\left\{\sigma_m^2\pmb{I}\right\}}_{m=1}^M,
\end{equation}
and
\begin{equation}
	\sigma_m^2 = \NF\cdot N_0 f_{\Delta,m}
\end{equation}
is the per-subcarrier noise variance for sub-band $m$, where $N_0$ is the white Gaussian noise power spectral density and $\NF$ is the noise figure. 

In this work, our goal is to investigate the impact of \ac{DMC} on multi-band sensing. Therefore, communication–sensing trade-offs in channel estimation are not incorporated into the model. We remark that when channel estimation is performed using payload data instead of pilots, the resulting estimation error is generally colored rather than white. Nevertheless, the conclusions of this study remain valid, since our model accommodates an arbitrary noise covariance matrix.

\section{Fundamental Limits and Empirical Metrics}
\label{sec:fundamental-limits-and-empirical-metrics}
Our problem definition for this work is \oscar{as follows. Given the observed background \ac{DMC}-affected multi-band \ac{MIMO} channel \ac{TF} $\pmb{h}$ at the \ac{Rx}, the estimator should jointly determine the number of paths $K$ and estimates their geometric parameters and complex gains. The parameters of interest are $\{\tau_k,\phi_k,\theta_k\}_{k=1}^{K}$, whereas $\{\Re g_{k,m},\Im g_{k,m}\}_{k,m = 1,1}^{K,M}$ are estimated as nuisance parameters. The following \ac{CRB} analysis is computed from the complete parameter vector containing both sets of parameters.} In this paper, we assume the covariance matrices of the background \ac{DMC} and white Gaussian noise are known.

The \ac{CRB}-based fundamental limits and the empirical metrics of the system discussed in this section are crucial for drawing intuitive and novel insights on multi-band sensing by capturing the interplay between \ac{SC}, \ac{DMC}, and noise effects simultaneously. They are also helpful for designing estimation algorithms by signaling the performance bounds.

\subsection{Cram\'er-Rao Bound (CRB)}
\label{subsec:cramer-rao-bound}
Assuming independence between the background \ac{DMC} and Gaussian noise, the distribution of $\pmb{h}$ can be modeled as 
\begin{equation}
	\pmb{h} \sim \mathcal{CN}\left(\pmb{s},\pmb{M} \right)
\end{equation} 
where
\begin{equation}
\pmb{M} = \blockdiag\left\{\pmb{M}_m\right\}_{m=1}^M = \blockdiag\left\{\pmb{R}_m+\pmb{S}_m\right\}_{m=1}^M
\end{equation}
is known in the context of this paper.
Hence, exploiting the block-diagonal structure of $\pmb{M}$, following the general complex Gaussian \ac{FIM} structure in \cite{kay1993estimation}, the \ac{FIM} is constructed as follows
\begin{equation}
\label{FIM}
	\pmb{F} = \sum_{m=1}^M \pmb{F}_m = \sum_{m=1}^M 2 \Re\left({\pmb{D}_m^H\pmb{M}_m^{-1}\pmb{D}_m}\right),
\end{equation}
where $\pmb{D}_m$ is the Jacobian formed by the partial derivatives of the mean vector $\pmb{s}$ (the \ac{SC}) with respect to the sensing parameters on each sub-band. 
The \ac{CRB} of the $i^{\text{th}}$ parameter can hence be obtained by $\CRB_{i} = \left[\pmb{F}^{-1}\right]_{i,i}$. 

We now derive expressions for the Jacobian matrix $\pmb{D}$ of the \ac{FIM}. 
We need
\begin{equation}
	\pmb{D}_m = 
	\cat_2\left\{
		\pmb{D}_m^{\tau},
		\pmb{D}_m^{\phi},
		\pmb{D}_m^{\theta},
		\pmb{D}_m^{\Re g},
		\pmb{D}_m^{\Im g}
	\right\}.
\end{equation}
\oscar{Since the \acp{CRB} for the geometric parameters are obtained after inverting the \ac{FIM} with the complete parameter vector, they account for the uncertainty introduced by estimating the nuisance path gains.}

For $\pmb{s}$ with respect to the geometric parameters, we have
\begin{equation}
\label{DsToA}
	\pmb{D}_m^{\tau} = 
	\cat_2\left\{
	g_{k,m} \frac{\partial}{\partial \tau_k}
	\pmb{a}_m (\tau_k,\phi_k,\theta_k)
	\right\}_{k=1}^K,
\end{equation}
and $\pmb{D}_{\phi}$ and $\pmb{D}_{\theta}$ follow similar structures from \eqref{DsToA} by changing $\tau$ to $\phi$ and $\theta$. Due to the Kronecker structure, we only need to know the following:
\begin{equation}
\label{partialaFToA}
	\frac{\partial \pmb{a}^{\tt{F}}_m}{\partial \tau_k} = 
	\pmb{a}^{\tt{F}}_m \odot 
	\left[
		-jn\omega_{\Delta,m}
	\right]_{n=0}^{N-1},
\end{equation}
\begin{equation}
\label{partialaTxAoD}
	\frac{\partial \pmb{a}^{\tt{T}}_m}{\partial \phi_k} = 
	\pmb{a}^{\tt{T}}_m \odot
	\left[
		-j\omega_m  (d^{\tt{T}}/c) \ell \cos \phi_k
	\right]_{\ell=0}^{L^{\tt{T}}-1},
\end{equation}
\begin{equation}
\label{partialaRxAoA}
	\frac{\partial \pmb{a}^{\tt{R}}_m}{\partial \theta_k} = 
	\pmb{a}^{\tt{R}}_m \odot
	\left[
		-j\omega_m  (d^{\tt{R}}/c) \ell \cos \theta_k
	\right]_{\ell=0}^{L^{\tt{R}}-1},
\end{equation}
Then, for $\pmb{s}$ with respect to the gains, we have
\begin{equation}
	\pmb{D}_m^{\Re g}
	=
	\cat_2\left\{
	\left[
	\pmb{0}_{\times (m-1)}, 
		\pmb{a}_m(\tau_k,\phi_k,\theta_k), 
	\pmb{0}_{\times (M-m)}
	\right]
	\right\}_{k=1}^K,
\end{equation}
\begin{equation}
	\pmb{D}_m^{\Im g}
	=
	\cat_2\left\{
	\left[
	\pmb{0}_{\times (m-1)}, 
		j\pmb{a}_m(\tau_k,\phi_k,\theta_k), 
	\pmb{0}_{\times (M-m)}
	\right]
	\right\}_{k=1}^K.
\end{equation}
\oscar{Because the sub-bands are narrow and widely separated, $g_{k,m}$ is assumed constant within each sub-band $m$, while a separate complex gain is estimated for every sub-band. The frequency dependence of $\gamma_{k,m}$ is therefore absorbed into the independent nuisance parameters $g_{k,m}$. Moreover, although the physical gain model depends on propagation distance, our estimation algorithm and thus our \ac{CRB} treat the gains and delays as independent unknowns.}

\subsection{\oscar{The Misspecified CRB}}

\oscar{Although the covariance matrices $\{\pmb{M}_m\}_{m=1}^{M}$ are assumed to be perfectly known above, they must be estimated in practice. We therefore investigate the sensitivity of the \ac{CRB} to errors in the \ac{DMC} parameters $\alpha$ and $\{\beta_m\}_{m=1}^{M}$ using the classical \ac{MCRB}, interpreted as the sandwich covariance associated with the misspecified maximum-likelihood estimator \cite{RichmondHorowitz2015,KhatibEtAl2026}. Note that $\alpha$ and $\{\beta_m\}_{m=1}^{M}$ are never in the parameter vector for calculating the \ac{FIM}. Instead, they are assumed known in the current model. Thus, to study the impact due to errors in $\alpha$ and $\{\beta_m\}_{m=1}^{M}$ on the \ac{CRB}, which are model assumption errors, the \ac{MCRB} is the appropriate tool.}

\oscar{Let $\pmb{M}_m=\pmb{M}_m(\alpha,\beta_m)$ denote the true covariance and $\widehat{\pmb{M}}_m=\pmb{M}_m(\widehat{\alpha},\widehat{\beta}_m)$ the covariance assumed by the estimator. For the covariance-only mismatch considered here, the mean vector $\pmb{s}_m$ and its Jacobian $\pmb{D}_m$ remain correctly specified, while both covariance matrices are treated as fixed with respect to the sensing parameters. The corresponding Hessian- and gradient-based information matrices are}
\begin{equation}
\oscar{%
\label{eq:misspecified-information-hess}
\pmb{A}
=
\sum_{m=1}^{M}
2\Re\!\left\{
\pmb{D}_m^{H}
\widehat{\pmb{M}}_m^{-1}
\pmb{D}_m
\right\},
}%
\end{equation}
\begin{equation}
\oscar{%
\label{eq:misspecified-information-score}
\pmb{B}
=
\sum_{m=1}^{M}
2\Re\!\left\{
\pmb{D}_m^{H}
\widehat{\pmb{M}}_m^{-1}
\pmb{M}_m
\widehat{\pmb{M}}_m^{-1}
\pmb{D}_m
\right\}.
}%
\end{equation}
\oscar{The misspecified bound is then \cite{RichmondHorowitz2015,KhatibEtAl2026}}
\begin{equation}
\oscar{%
\MCRB_i
=
\left[
\pmb{A}^{-1}\pmb{B}\pmb{A}^{-1}
\right]_{i,i}.
\label{eq:mcrb}
}%
\end{equation}
\oscar{When $\widehat{\pmb{M}}_m=\pmb{M}_m$ for every sub-band, $\pmb{A}=\pmb{B}=\pmb{F}$ and \eqref{eq:mcrb} reduces to the conventional $\CRB_i$. We quantify the resulting sensitivity through the normalized degradation of parameter $i$,}
\begin{equation}
\oscar{%
\Delta\CRB_i
=
10\log_{10}
\left(
\frac{\MCRB_i}
{\CRB_i}
\right).
\label{eq:normalized-mcrb}
}%
\end{equation}

\subsection{The Approximate CRB}
In Section \ref{sec:proposed-multi-band-estimation-algorithm}, we propose a multi-band estimator based on \ac{CRB}-weighted combining. To assess the performance of this estimator, we need to derive its \ac{CRB}, which is an \oscar{upper-bound} approximation to the exact joint \ac{CRB} derived in the previous subsection. This approximate \ac{CRB} also plays a crucial role in evaluating the performance loss of the proposed estimator compared to the ideal joint multi-band estimator. The approximate \ac{CRB} takes the form
\begin{equation}
\label{approxcrb}
	\CRB^{\tt{approx}}_{\psi_k} = \left(\sum_{m=1}^M \CRB^{-1}_{\psi_k,m}\right)^{-1}
\end{equation}
 (see Appendix \ref{appendix:derivations-of-the-minimal-variance-linear-unbiased-estimator} for detailed derivations). 
 
 This approximation is equivalent to ignoring the cross terms in the \acp{FIM} $\pmb{F}_m, \forall m$. Thus, it is obvious that if the paths are less coupled, the approximate \ac{CRB} better approximates the exact \ac{CRB}.
 Also, since all \ac{CRB} terms are positive, we can notice that
 \begin{equation}
 	\sum_{m'=1}^M \CRB^{-1}_{\psi_k,m'} \geq
 	\CRB^{-1}_{\psi_k,m}, \forall m.
 \end{equation}
Inverting both sides and using the definition in \eqref{approxcrb}, we obtain
\begin{equation}
	\CRB^{\tt{approx}}_{\psi_k} \leq
 	\CRB_{\psi_k,m}, \forall m,
 \end{equation}
 suggesting that the approximate multi-band \ac{CRB} discussed in this subsection serves as a lower bound to the exact single-band \acp{CRB} for any sub-band.
 
\subsection{Estimation Signal-to-Noise Ratio (ESNR)}

The \ac{ESNR} can be thought of as a measure of the certainty of an estimated path. One crucial issue that needs to be addressed when estimating the \ac{SC} parameters is how the algorithm should determine whether a peak is \ac{SC} or part of the \ac{DMC} profile. Another similar issue is whether the algorithm can trust estimates from a particular sub-band when combining across sub-bands. Both of these issues could be addressed by using \ac{ESNR}.

For a true path with gain $g_{k,m}$, the \ac{ESNR} $\mu_{k,m}$ is defined as
\begin{equation}
	\mu_{k,m} = \frac{\lvert g_{k,m} \rvert^2}{\CRB_{\lvert g_{k,m} \rvert}},
\end{equation}
where $\CRB_{\lvert g_{k,m} \rvert}$ can be obtained from $\CRB_{\Re g_{k,m}}$ and $\CRB_{\Im g_{k,m}}$ through a Jacobian coordinate transformation on the \ac{FIM}. We can see that if the \ac{ESNR} is low, the variance of our estimate will be so large compared to the true value that it is no longer informative for estimation. In our work, the \ac{ESNR} plays an instrumental role in multi-band sensing, serving as the threshold for \oscar{selecting which path estimates to retain for multi-band combination}.

\subsection{Bistatic Geometric Gain (BGG)}

To investigate the combined effect of both \ac{DMC} and noise on the \ac{SC}, it is meaningful to examine a metric that captures their joint spatial interplay while preserving the geometric intuition commonly used in ISAC and radar analysis. In classical bistatic radar formulations, performance metrics such as received power or detection \ac{SNR} depend on the product of \ac{Tx}-scatterer and scatterer-\ac{Rx} distances, naturally giving rise to \textit{Cassini ovals} as contours of equal sensing performance \cite{willis2005bistatic,merrill2008radar}. These ovals provide an intuitive visualization of geometry-induced performance variations under free-space and noise-limited assumptions.

In our more realistic ISAC case, however, sensing performance is not governed solely by the \ac{SC} and thermal noise. The presence of \ac{DMC} introduces a spatially structured interference term whose power depends on both delay- and angular-domain profiles, fundamentally altering the spatial behavior of estimation performance. Interpretations from classical Cassini ovals, therefore, become insufficient, as they neglect the frequency-dependent and geometry-dependent impact of DMC. This consequence motivates the introduction of a metric that explicitly accounts for the combined contributions of \ac{SC}, \ac{DMC}, and noise on sensing performance.

To this end, we define the \ac{BGG} metric for the $m^{\textrm{th}}$ sub-band as
\begin{equation}
	\Gamma_m(\pmb{p}) = 
	\frac{G^{\tt{SC}}_m(\pmb{p})}{G^{\tt{DMC}}_m(\pmb{p})+\sigma_m^2},
\end{equation}
where $\pmb{p}$ denotes the Cartesian coordinates of a 2D location. The numerator
\begin{equation}
	G^{\tt{SC}}_m(\pmb{p}) = 
	\frac
	{P^{\tt{T}}f_{\Delta,m}}
	{(4\pi)^2(c\tau^{\tt{D}}(\pmb{p}))^2(c\tau^{\tt{A}}(\pmb{p}))^2}
\end{equation}
corresponds to the geometry-induced power of the \ac{SC}, which depends on the product of bistatic path lengths. The denominator
\begin{equation}
	G^{\tt{DMC}}_m(\pmb{p}) = 
		\frac{P^{\tt{DS}}_m(\tau(\pmb{p}),\tau_1)\cdot 
		P^{\tt{AS}}_m(\phi(\pmb{p}),\phi_1)\cdot 
		P^{\tt{AS}}_m(\theta(\pmb{p}),\theta_1)}{\magn{\gamma_{1,m}}^2},
\end{equation}
together with the sub-band noise variance $\sigma_m^2$,
captures the spatially varying, geometry-induced contribution of \ac{DMC}. 

By construction, the \ac{BGG} metric isolates the geometry-dependent interaction between \ac{SC}, \ac{DMC}, and noise by factoring out path-dependent coefficients $\gamma_{k,m}$ and the impact of the sub-band dependent antenna directivity patterns $u^{\tt{T}}_m(\phi)$ and $u^{\tt{R}}_m(\theta)$ on the angular distribution of \ac{DMC}. Its level sets can therefore be interpreted as \textit{modified Cassini ovals}, where equal values of $\Gamma_m$ correspond to locations exhibiting comparable \ac{SINR}-like conditions for estimation. When $G^{\tt{DMC}}_m(\pmb{p})$ = 0, i.e., in the absence of \ac{DMC} interference, the modified Cassini ovals naturally coincide with the conventional Cassini ovals. As such, the \ac{BGG} metric provides a physically meaningful and interpretable proxy for sensing performance in \ac{DMC}-affected multi-band ISAC scenarios.

\subsection{Probabilities of Detection and False Alarm}

One of the main advantages of our proposed multi-band estimation algorithm outlined in Section \ref{sec:proposed-multi-band-estimation-algorithm} is its ability to resolve the angular ambiguities caused by the grating lobes effect. To quantitatively characterize this behavior, we introduce the \ac{PD} and \ac{PFA} metrics and investigate the resulting \ac{PD}-\ac{PFA} curves. 

Let $\mathcal{H}_K$ denote the hypothesis that the environment contains $K$ paths (including the \ac{LoS} path), and let
$\hat{\mathcal X}\subseteq\mathbb{R}^3$ \oscar{denote} the set containing the geometric parameters of all estimated paths returned by the algorithm, where each element is
$\hat{\pmb{x}}=[\hat{\tau},\hat{\phi},\hat{\theta}]$.
For the $k^{\textrm{th}}$ true path with geometric parameters
$\pmb{x}_k=[\tau_k,\phi_k,\theta_k]$, we define its region of detection as
\begin{equation}
\mathcal{D}_k
=
\left\{
\hat{\pmb{x}}\in\mathbb{R}^3
\,:\,
\norm{\mathcal{T}\left(\hat{\pmb{x}}\right)-\mathcal{T}\left(\pmb{x}_k\right)}_2
\leq R
\right\},
\end{equation}
where $R$ is the detection radius and $\mathcal{T}(\cdot)$ is a resolution-aware normalization mapping explained in detail in the later paragraphs.
We then define the \ac{PD} and \ac{PFA} as
\begin{equation}
\PD
=
\Pr\left(
\forall k\in\{1,2,\dots,K\},\,
\left|\hat{\mathcal X}\cap \mathcal{D}_k\right|>0
\,\middle|\, \mathcal{H}_K
\right),
\end{equation}
\begin{equation}
\PFA
=
\Pr\left(
\exists\,\hat{\pmb{x}}\in\hat{\mathcal X}\ \text{s.t.}\ 
\hat{\pmb{x}}\notin \bigcup_{k=1}^{K}\mathcal{D}_k
\,\middle|\, \mathcal{H}_K
\right).
\end{equation}

The normalization mapping \oscar{$\mathcal{T}: \mathbb{R}\times \left(-90^\circ,90^\circ\right)^2 \to \mathbb{R}\times\left(-\frac{L^{\tt{T}}}{2}, \frac{L^{\tt{T}}}{2}\right) \times \left(-\frac{L^{\tt{R}}}{2}, \frac{L^{\tt{R}}}{2}\right)$} is defined as
\begin{equation}
\oscar{%
\label{trans}
\mathcal{T}\left(
   \left[ 
\tau,
\phi,
\theta
\right]\right)
= 
\left[N f_{\Delta,0} \tau,\frac{L^{\tt{T}}}{2}\sin \phi,\frac{L^{\tt{R}}}{2}\sin \theta\right].
}%
\end{equation}
It is designed to compare delay/angle values in a dimensionless, resolution-aware coordinate system that matches the natural phase variables of the \ac{OFDM}-\ac{MIMO} steering vectors. 

Specifically, we acknowledge a well-known relation that the delay resolution scales with $1/(N f_{\Delta})$, where $N f_{\Delta}$ is the occupied bandwidth. Since the goal is to normalize the delays and angles for comparison across sub-bands, we choose a reference subcarrier spacing $f_{\Delta,0}$ to be $f_{\Delta,1}$, the subcarrier spacing of the lowest sub-band. Dividing the delay value $\tau$ by the corresponding resolution yields the normalized delay coordinate $N f_{\Delta,0} \tau$. To extend this approach to the angle domain, we observe a structural similarity between the frequency and space-domain steering vectors. While delay induces a linear phase progression across subcarriers through the term $e^{-j2\pi n f_{\Delta}\tau}$, angles induce a linear phase progression across \ac{ULA} array elements through the spatial frequency term $e^{-j2\pi (d/\lambda)\ell \sin(\cdot)}$. Consequently, angular resolution scales with $1/(L(d/\lambda))$ if the angle is transformed by $\sin(\cdot)$. By selecting a reference wavelength $\lambda_0$ such that $\lambda_0 = 2 d$, the normalized angular coordinates \oscar{$L^{\tt{T}}\sin \phi/2$ and $L^{\tt{R}}\sin \theta/2$} are produced. As a result, Euclidean distances after $\mathcal{T}$ reflect discrepancies in approximately comparable resolution units across delay and angle, preventing the matching metric from being dominated by heterogeneous physical units and better aligning the association cost with the steering-vector sensitivity.

\section{Proposed Multi-Band Estimation Algorithm}
\label{sec:proposed-multi-band-estimation-algorithm}
In this section, we propose a scalable multi-band estimator based on \oscar{non-coherent, parameter-level fusion of single-band estimates}. \oscar{This approach has several advantages compared to coherent combination examples in works such as \cite{Pegoraro2026MultibandFR3} and \cite{Pegoraro2024HiSAC}. First, since we focus on investigating the gains of multi-band processing, non-coherent combination provides a practical solution for tractability with minimum performance loss and less complexity. Second, non-coherent combination cleans the measurements in each sub-band separately, thus preventing the occurrence of grating lobe artifacts \textit{in the range domain} \cite{Pegoraro2026MultibandFR3}.} 

\oscar{Nonetheless, it should also be noted that because the measurements are processed separately before fusion, the method does not synthesize the frequency span between sub-bands and therefore does not improve the nominal delay resolution or the ability to separate paths that are unresolved within every constituent sub-band. Its expected benefits are instead improved estimation accuracy for the already resolved paths and suppression of false alarms due to angular ambiguity through cross-sub-band association.}

For the single-band estimators, we choose the well-known state-of-the-art estimation algorithm proposed in \cite{rimaxthesis}, based on the iterative \ac{LM} approach. This estimator is known to achieve the \ac{CRB}.

\subsection{Overview of the Algorithm}

To easily accommodate multi-band operation, we wish to perform a linear combination of the estimated geometric parameter values (\acp{ToA}, \acp{AoA}, and \acp{AoD}) from the single-band estimators. When the estimators for every sub-band are efficient (i.e., they approach their own \ac{CRB}), we can obtain the minimal variance linear unbiased estimator by designing weights (for any sub-band $m$ where some geometric parameter $\psi_k$ is estimated) as \cite{kay1993estimation}
\begin{equation}
\label{weights}
	w_{\hat{\psi}_k,m} = \frac{\CRB^{-1}_{\hat{\psi}_k,m}}
	{\sum_{\forall m \text{ detected}} \CRB^{-1}_{\hat{\psi}_k,m}}
\end{equation}
(see Appendix \ref{appendix:derivations-of-the-minimal-variance-linear-unbiased-estimator} for detailed derivations).

Since our antenna spacings $d^{\tt{T}}$ and $d^{\tt{R}}$ are fixed while multiple frequencies are used, we could observe \textit{grating lobes} (i.e. spatial aliasing in angular domains) for certain \acp{AoA} or \acp{AoD} when $d > \lambda_m / 2$. Take \ac{AoD} as an example. For the $i^{\textrm{th}}$ real angle $\hat{\phi}_i$ estimated from a single-band estimator with wavelength $\lambda$,  aliased angles $\hat{\phi}^r_i$ should appear at \cite{mailloux2017phased}
\begin{equation}
\label{aliasedangles}
	\hat{\phi}^r_i = \sin^{-1} \big(\sin \hat{\phi}_i + r \lambda / d^{\tt{T}}\big),
\end{equation}
where $r \in \{ \pm 1, \pm 2, \dots\}$ can be any value that makes $\sin^{-1}(\cdot)$ take values in $[-1,1]$. We can also denote $\hat{\phi}_i$ by $\hat{\phi}^0_i$. 
Hence, for each estimated real path, we consider all its aliased versions as a \textit{path group}. Across the three records, the same path group is indexed by the same index $p$. Also, note that a perfect scheme for resolving ambiguity is beyond the scope of this paper.

The algorithm iterates over every sub-band $m$, and it dynamically keeps and updates, per valid iteration, three data containers of all estimated path groups:
\begin{itemize}
	\item \textbf{Estimates of Unambiguous Groups} $\mathcal{E}_U$: the latest weighted combined estimates of the unambiguous path groups ($p \in \mathcal{U}_m$). At each sub-band iteration $m$, the term corresponding to each geometric parameter is a scalar. For example, this record can contain the latest weighted combined \ac{AoD} estimate for path 2, denoted by $\hat{\phi}_2$.
	\item \textbf{Estimates of Ambiguous Groups} $\mathcal{E}_A$: the latest estimates of the ambiguous path groups. At each sub-band iteration $m$, the term corresponding to each geometric parameter is a set. For example, this record can contain the latest \ac{AoD} estimate for the ambiguous path group 3, denoted by $\hat{\phi}_3 = \big\{\hat{\phi}_3^r\big\}_{r \in \left\{0, \pm 1, \pm 2, \dots\right\}}$.
	\item \textbf{Record of Unambiguous Groups} $\mathcal{R}_U$: the record for the estimates per iteration $m$ of the unambiguous path groups ($p \in \mathcal{U}_m$).  At each sub-band iteration $m$, the term corresponding to each geometric parameter is a $1\times m$ vector. For example, this record can contain a vector with all past \ac{AoD} estimates for path 2, denoted by $\hat{\pmb{\phi}}_2 = \big[ \hat{\phi}_2^{(1)},\hat{\phi}_2^{(2)},\dots,\hat{\phi}_2^{(m)} \big]$.
\end{itemize}

\begin{figure}[!t]
\centering
\includegraphics[width=3.5in]{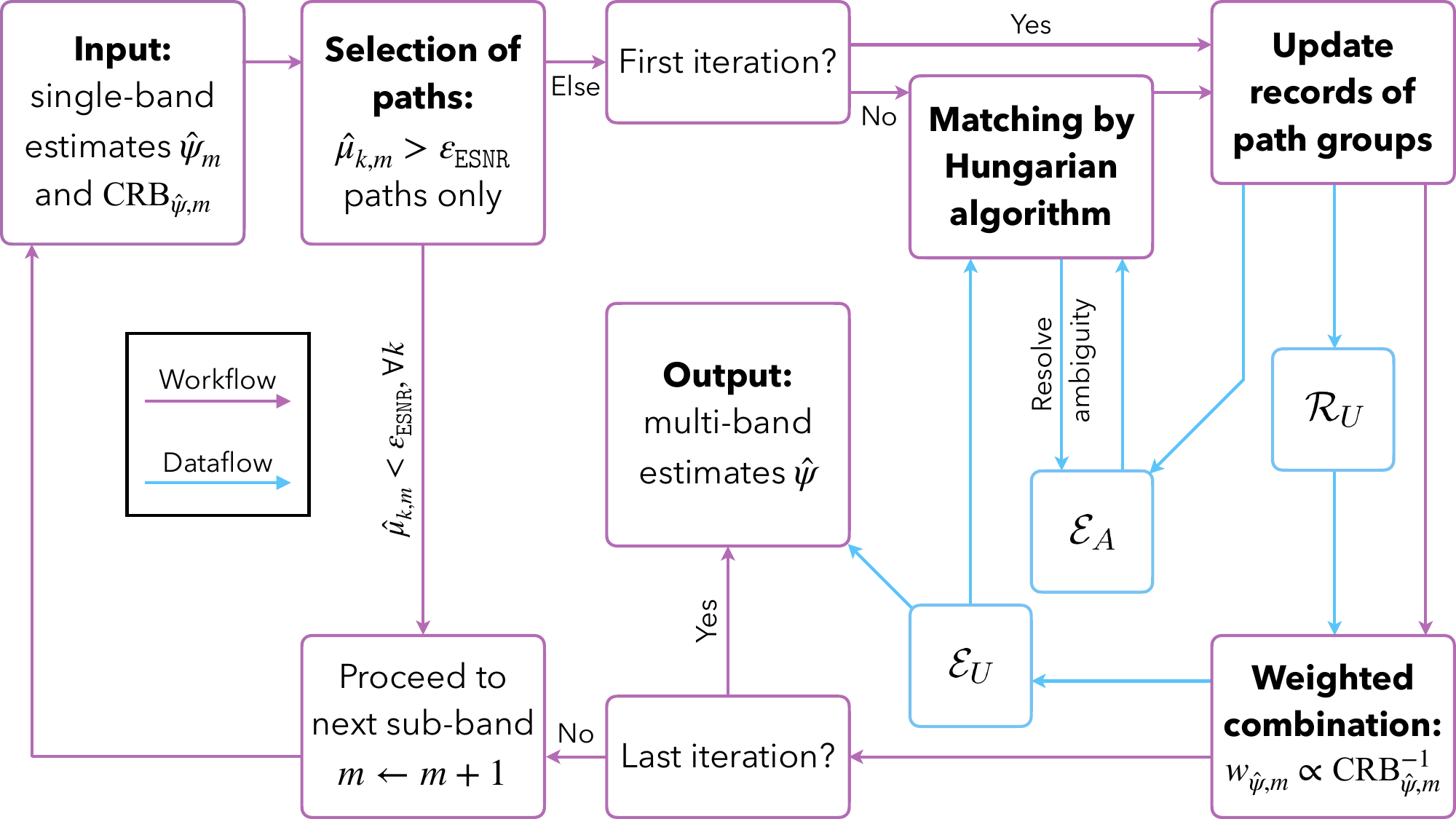}
\caption{\oscar{Flowchart of the algorithm. Each subsection that discusses the algorithm from \ref{subsec:selection-of-paths} to \ref{subsec:weighted-combination} corresponds to a box with bolded text.}}
\label{fig:flowchart}
\end{figure}

A summary of the algorithm is provided in the flowchart of Fig. \ref{fig:flowchart}, where $\hat{\psi}$ refers to the estimate for any geometric parameter $\psi$ (including \ac{ToA} $\tau$, \ac{AoD} $\phi$, or \ac{AoA} $\theta$). \oscar{Details of the key steps of the algorithm are discussed in the later subsections \ref{subsec:selection-of-paths} to \ref{subsec:weighted-combination}.}
 
\subsection{\oscar{Selection of Paths}}
\label{subsec:selection-of-paths}
Since \ac{ESNR} is a measure of how certain we can trust a path seen by a given sub-band, we use it for \oscar{path} selection. In the algorithm, the estimated values are used instead of the true values to form the estimated \ac{ESNR}
\begin{equation}
\hat{\mu}_{k,m} = \frac{\lvert \hat{g}_{k,m} \rvert^2}{\CRB_{\lvert \hat{g}_{k,m} \rvert}}.
\end{equation}
\oscar{Only paths with $\hat{\mu}_{k,m} > \varepsilon_{\tt{ESNR}}$ are retained by the multi-band path-selection stage, so if}
 $\hat{\mu}_{k,m} < \varepsilon_{\tt{ESNR}}, \forall k$, where $\varepsilon_{\tt{ESNR}}$ is the predefined \ac{ESNR} threshold, the \oscar{entire} sub-band is uncertain and should be \oscar{skipped} for combining. 
 
\subsection{Matching by Hungarian Algorithm}

To simultaneously associate estimates across sub-bands and eliminate the aliased estimated angles, we use the Hungarian matching algorithm.
In each valid iteration on sub-band $m$, we first perform the matching between the overall latest estimated groups (the terms in $\mathcal{E}_U$ and $\mathcal{E}_A$, whose index is $p$) and the newly estimated path groups from the single-band estimator of this sub-band (whose index is $i$). The inter-group matching cost $\left[{\pmb{C}}_{m}\right]_{p,i}$, which is the $(p,i)^\textrm{th}$ term of the Hungarian matching cost matrix ${\pmb{C}}_m$, is computed based the minimal $\ell_2$-distance between all ambiguity versions of the transformed geometric parameters of both groups as follows
\begin{equation}
\label{hungcost}
	\left[{\pmb{C}}_{m}\right]_{p,i} = \min\limits_{r_{1:4}}
    \norm{
    \mathcal{T}\left( \hat{\pmb{x}}_p^{r_1,r_2}\right) - 
    \mathcal{T}\left(\hat{\pmb{x}}_i^{r_3,r_4}\right)}_2,
\end{equation} 
\begin{equation}
	r^{\star}_{1:4} = \argmin\limits_{r_{1:4}}
    \norm{
    \mathcal{T}\left( \hat{\pmb{x}}_p^{r_1,r_2}\right) - 
    \mathcal{T}\left(\hat{\pmb{x}}_i^{r_3,r_4}\right)}_2,
\end{equation} 
where $\hat{\pmb{x}}_p^{r_1,r_2}$ denotes $[\hat{\tau}_p, \hat{\phi}^{r_1}_p, \hat{\theta}^{r_2}_p]$, $\hat{\pmb{x}}_i^{r_3,r_4}$ denotes $[\hat{\tau}_i, \hat{\phi}^{r_3}_i, \hat{\theta}^{r_4}_p]$, $r^{\star}_{1:4}$ stores the optimal ambiguity version indices for computing this cost, and $\mathcal{T}$ denotes the same resolution-aware normalization mapping defined in \eqref{trans}. Note that the Hungarian matching algorithm can optionally take a maximum cost input, $C_{\tt{max}}$, to prevent matching two path groups that are far apart. We also compute the prominence $\left[{\pmb{P}}_{m}\right]_{p,i}$ for the matching cost of each matched pair of groups $(p,i)$, which is defined as 
\begin{equation}
	\left[{\pmb{P}}_{m}\right]_{p,i} = \min\limits_{r_{1:4} \neq r^{\star}_{1:4}}
    \norm{
    \mathcal{T}\left( \hat{\pmb{x}}_p^{r_1,r_2}\right) - 
    \mathcal{T}\left(\hat{\pmb{x}}_i^{r_3,r_4}\right)}_2
    -
    \left[{\pmb{C}}_{m}\right]_{p,i}.
\end{equation}

The ambiguity of an ambiguous $p$-indexed overall latest group is resolved if both of the following conditions hold: 
\begin{itemize}
	\item A matching exists between it and some $i$-indexed newly estimated path group after running the Hungarian matching algorithm,
	\item The matched pair of path groups $(p,i)$ satisfies the minimum prominence condition $\left[{\pmb{P}}_{m}\right]_{p,i} > \varepsilon_{\tt{prom}}$, where $\varepsilon_{\tt{prom}}$ is the predefined prominence threshold.
\end{itemize}
We also know which aliased version should now be regarded as the true estimate of this previously ambiguous group. For example, we can now deduce that $\hat{\phi}_p^{(m_a)} = \hat{\phi}^{r^\star_1}_p$ and $\hat{\phi}_i^{(m)} = \hat{\phi}^{r_3}_i$, where $m_a$ is the iteration index of the sub-band where the ambiguity of the group is first observed.

\subsection{Update Records of Path Groups}
If some $i$-indexed newly estimated path groups are not matched to any path groups in $\mathcal{E}_U$ or $\mathcal{E}_A$, we call them \textit{new path groups} and assign to them new $p$-indices.
The ambiguous new path groups are directly appended to the end of $\mathcal{E}_A$, and the unambiguous new path groups are used to update $\mathcal{R}_U$ before generation of the latest $\mathcal{E}_U$ through weighted combination.
 We denote $\mathcal{U}_m$ as the set of new $p$'s which represent unambiguous path groups after these steps, whose number of elements is used to update $\hat{K}$, the estimated number of paths.
 
Now, we discuss how to update $\mathcal{R}_U$ by taking \ac{AoD} as an example. For each $p \in \mathcal{U}_m$, we consider updating as follows the row vectors for both the parameter estimates $\hat{\pmb{\phi}}_p$ and the corresponding (unnormalized) weight estimates ${\pmb{c}^{\hat{\phi}}_p}$:
\begin{equation}
\label{paramstore}
\begin{aligned}
	\big[\hat{\pmb{\phi}}_p\big]_m = 
	\begin{cases}
	\hat{\phi}_i & \textrm{if } p \textrm{ is matched to or converted from an }i,\\
	0 & \textrm{otherwise},
	\end{cases}
	\end{aligned}
\end{equation}
\begin{equation}
\label{paramstore2}
\begin{aligned}
	\big[\hat{\pmb{\phi}}_p\big]_{1:m-1} = 
	\begin{cases}
	\pmb{0}_{1 \times m-1} & \textrm{if } p \textrm{ is converted from an }i,\\
	\big[\pmb{0},\hat{\phi}^{(m_a)}_p,\pmb{0}\big] & \textrm{if } p \textrm{ was ambiguous at }m_a, \\
	\textrm{inherited} &\textrm{otherwise}. \\
	\end{cases}
	\end{aligned}
\end{equation}
The schemes for ${\pmb{c}^{\hat{\phi}}_p}$ are similar, where only replacing $\hat{\phi}_i$ and $\hat{\phi}_p$ by $\CRB^{-1}_{\hat{\phi}_i}$ and $\CRB^{-1}_{\hat{\phi}_p}$ are needed. Additionally, similar updating schemes are applied to the other sensing parameters (except for $g$, for which we do not calculate the \ac{CRB}-based weights vector).

\subsection{Weighted Combination}
\label{subsec:weighted-combination}
 Using the data in $\mathcal{R}_U$, we then update $\mathcal{E}_U$ for all $p\in\mathcal{U}_m$ by weighted combination across all previous sub-bands through
\begin{equation}
\label{fusion}
\hat{\phi}_p =
\frac{
\hat{\pmb{\phi}}_p \big({\pmb{c}^{\hat{\phi}}_p}\big) ^T
}{
{\pmb{c}^{\hat{\phi}}_p} \pmb{1}_{m\times 1}
},
\end{equation}
and the combination schemes of $\hat{\tau}_p, \hat{\theta}_p$ follow similar forms. If $p \notin \mathcal{U}_m$, we only store them in $\mathcal{E}_A$ for later matching, as we currently do not know which aliased version gives the real estimate. 

It is evident that even after we have proceeded to the last sub-band $M$, there may still be some remaining ambiguous path groups in $\mathcal{E}_A$. At this stage, we have no choice but to ignore them. So, for every $p \in \mathcal{U}_{M}$, we now have the final combined estimates of the geometric parameters from $\mathcal{E}_U$. For the gain parameters, we report the corresponding $\hat{g}_{p,m}$'s using the $\mathcal{R}_U$ record. Finally, we re-index the $p$-indices by $k$ and pass them to the algorithm's output. The final estimated number of paths, $\hat{K}$, is hence given by $\hat{K} = \magn{\mathcal{U}_M}$.

\subsection{\oscar{Computational Complexity Analysis}}
\subsubsection{\oscar{Single-Band Estimator}}

\begin{table}[!t]
\renewcommand{\arraystretch}{1.3}
\renewcommand{\tabularxcolumn}[1]{m{#1}}
\caption{\oscar{Computational Complexity of the Single-Band Estimator}}
\label{table:complexity_sb}
\centering
\newcolumntype{C}{>{\centering\arraybackslash}X}

\begin{tabularx}{\columnwidth}{|C|C|}
\hline
\rowcolor{gray10pct}
\textbf{Operation} & \textbf{Complexity Upper Bound}\\
\hline

$\pmb{R}^{\tt F}_m$ inversion
&
$\mathcal{O}(N^3)$
\\
\hline

Global search
&
$\mathcal{O}\left(K_{\tt max}D\log N\right)$
\\
\hline

Component matrices of the Jacobian
&
$\mathcal{O}\left(D K_{\tt max}^{2}\right)$
\\
\hline

Covariance weighting
&
$\mathcal{O}\left(
NDK_{\tt max}^{2}
\right)$
\\
\hline

Weighted gain Gram matrix
&
$\mathcal{O}\left(
DK_{\tt max}^{3}
\right)$
\\
\hline

Path weights estimation using Gram matrix
&
$\mathcal{O}\left(
K_{\tt max}^{4}
\right)$
\\
\hline

Approximate Hessian / \ac{FIM}
&
$\mathcal{O}
\left(
N^{2}K_{\tt max}^{2}
+
EK_{\tt max}^{3}
\right)$
\\
\hline

Parameter update
&
$\mathcal{O}\left(
K_{\tt max}^{4}
\right)$
\\
\hline

Score function and strict maximization check
&
$\mathcal{O}\left(
NDK_{\tt max}+DK_{\tt max}^{2}
\right)$
\\
\hline

\ac{CRB} computation
&
$\mathcal{O}\left(
N^{2}K_{\tt max}^{2}
+
EK_{\tt max}^{3}
+
K_{\tt max}^{4}
\right)$
\\
\hline

Conversion to physical delays and angles
&
$\mathcal{O}\left(
K_{\tt max}
\right)$
\\
\hline

\end{tabularx}
\end{table}

\oscar{The estimator repeats the following process until the number of paths reaches a predefined $K_{\tt max}$. First, it finds a new path using a \ac{FFT}-based global search. Second, it adds that path to the existing path list and refines the entire list using the \ac{LM} method outlined in \cite{rimaxthesis}. Third, it computes the \ac{CRB} and the \ac{ESNR}. The resulting estimates and \ac{ESNR} values are then passed to the multi-band path-selection stage. The single-band estimator exploits
$\pmb{M}_m^{-1}
=\pmb{I}_{L^{\tt T}L^{\tt R}}\otimes
(\pmb{R}^{\tt F}_m)^{-1}$ and, therefore, only inverts the
$N\times N$ frequency covariance matrix. Table~\ref{table:complexity_sb}
lists the worst case complexity of each step with respect to the system parameters $M$, $N$, $L^{\tt T}$, $L^{\tt R}$, and $K$ ($K_{\tt max}$ is chosen relative to $K$), where $D=NL^{\tt T}L^{\tt R}$ and $E = N+L^{\tt T}+L^{\tt R}$. Note that we treat the maximum number of \ac{LM} iterations as a fixed implementation constant, and we also treat the upper bounds of the numbers of \ac{AoD} and \ac{AoA} ambiguity candidates as constants unrelated to the system parameters of interest. Therefore, retaining the dominant terms gives}
\begin{equation}
\oscar{
\label{eq:complexity_sb}
T^{\tt SB}
=
\mathcal{O}\left(
N^{3}
+
K_{\tt max}D\log N
\\
+
NDK_{\tt max}^{2}
\\
+
DK_{\tt max}^{3}
+
K_{\tt max}^{4}
\right).
}
\end{equation}

\subsubsection{\oscar{Multi-Band Steps}}

\begin{table}[!t]
\renewcommand{\arraystretch}{1.3}
\renewcommand{\tabularxcolumn}[1]{m{#1}}
\caption{\oscar{Computational Complexity of the Overall Estimator}}
\label{table:complexity_overall}
\centering
\newcolumntype{C}{>{\centering\arraybackslash}X}

\begin{tabularx}{\columnwidth}{|C|C|}
\hline
\rowcolor{gray10pct}
\textbf{Operation} & \textbf{Complexity Upper Bound}\\
\hline

Single-band estimation
&
$\mathcal{O}\!\left(MT^{\tt SB}\right)$
\\
\hline

Selection of paths
&
$\mathcal{O}\!\left(MK_{\tt max}\right)$
\\
\hline

Matching-cost evaluation
&
$\mathcal{O}\!\left(
MK_{\tt max}^{2}
\right)$
\\
\hline

Hungarian assignment
&
$\mathcal{O}\!\left(MK_{\tt max}^{3}\right)$
\\
\hline

Record update and fusion
&
$\mathcal{O}\!\left(M^{2}K_{\tt max}\right)$
\\
\hline
\end{tabularx}
\end{table}

\oscar{The multi-band stage performs path selection based on \ac{ESNR}, ambiguity-aware
Hungarian matching, and CRB-weighted fusion. Table~\ref{table:complexity_overall} lists the worst case complexity of each multi-band step. Retaining the dominant terms gives the total complexity 
\begin{equation}
\label{eq:complexity_overall}
\begin{split}
T^{\tt tot}
  &= \mathcal{O}\left(
       MT^{\tt SB}
       + MK_{\tt max}^{3}
       + M^{2}K_{\tt max}
     \right) \\
  &= \mathcal{O}\big(
       MN^3
       + MK_{\tt max}D\log N
       + MNDK_{\tt max}^{2} \\
  &\qquad
       + MDK_{\tt max}^{3}
       + MK_{\tt max}^{4}
       + M^2K_{\tt max}
     \big).
\end{split}
\end{equation}
Thus, the complexity is generally dominated by the independent
single-band estimators.}

\subsubsection{\oscar{Comparison with Coherent Estimators}}

\oscar{For a fair comparison, consider a coherent estimator using the
same FFT-based initialization, sequential addition of estimated paths, fixed
number of LM iterations, and the same covariance structure $\pmb{M}_m^{-1}
=\pmb{I}_{L^{\tt T}L^{\tt R}}\otimes
(\pmb{R}^{\tt F}_m)^{-1}$ as the proposed
estimator. 
Hence, covariance inversion and initialization have the same
$\mathcal{O}(MN^3)$ and
$\mathcal{O}(MK_{\tt max}D\log N)$ costs in both approaches.}

\oscar{At the stage containing $k$ paths, the coherent parameter vector
contains $3k$ common geometric parameters and $2Mk$ real-valued
sub-band gain parameters. Using the same Kronecker covariance operations,
the covariance weighting and information-matrix construction over all
bands cost $\mathcal{O}(MNDk+MDk^2)$. A dense solution of the joint \ac{LM}
parameter update equation costs $\mathcal{O}\left((3k+2Mk)^3\right)$. Using the same upper-bounding approach, in which we count $K$ instead of $k$ paths for each \ac{LM} stage, gives}
\begin{equation}
\oscar{
\label{eq:complexity_coherent}
\begin{split}
T^{\tt coh}
=\mathcal{O}\big(
&MN^3
+MK_{\tt max}D\log N
+MNDK_{\tt max}^{2}
\\
&+MDK_{\tt max}^{3}
+(3+2M)^3K_{\tt max}^{4}
\big).
\end{split}
}
\end{equation}

\oscar{Comparing with \eqref{eq:complexity_overall},
the difference lies in the following term: dense coherent processing
requires
$\mathcal{O}((3+2M)^3K_{\tt max}^{4})
=\mathcal{O}(M^3K_{\tt max}^{4})$,
whereas the proposed estimator requires
$\mathcal{O}(MK_{\tt max}^{4})$ for the independent parameter
solutions and $\mathcal{O}(M^2K_{\tt max})$ for fusion. Therefore, the
dense coherent solve is cubic in $M$, while the additional fusion cost
of the proposed estimator is at most quadratic in $M$.}

\oscar{Nonetheless, we admit that a structure-aware coherent solver may remove the dense cubic
penalty and achieve a comparable or lower complexity. However, the proposed
estimator instead provides parallelism: all $M$
single-band estimators can be executed independently and concurrently. A structure-aware
coherent estimator can parallelize some sub-band-wise operations, but it
must combine their results during every refinement iteration to update
the shared geometric parameters.}

\section{Numerical Results}
\label{sec:numerical-results}
\begin{table}[!t]
\renewcommand{\arraystretch}{1.3}
\caption{\oscar{Numerical Parameters for Figs. \ref{fig:crb-pt-var_alpha} to \ref{fig:simulation_PD_PFA-var_thresh}}}
\label{table:numerical-parameters}
\centering

\begin{tabularx}{\columnwidth}{|>{\centering\arraybackslash}m{0.2\columnwidth}|>{\centering\arraybackslash}X|}
\hline
\rowcolor{gray10pct} \multicolumn{2}{|c|}{\bfseries System Parameters} \\
\hline
$N$ & 128 \\
\hline
$d^{\tt{T}}$, $d^{\tt{R}}$ & 0.02 m \\
\hline
$f_m$ & $\{8.75,10,15,21.7\} \text{ GHz}$ \\
\hline
$f_{\Delta,m}$ & 1 MHz, $\forall m$ \\
\hline 
$\tilde{\beta}_m$ & $\{0.5,1,1.25,1.5\}$ \\
\hline
$\pmb{R}_m^{\tt{T}}$, $\pmb{R}_m^{\tt{R}}$ & $\pmb{I}, \forall m$ (DMC angular spread not modeled) \\
\hline
$N_0$ & -174 dBm/Hz\\
\hline
$\NF$ & 7 dB \\
\hline
\rowcolor{gray10pct}
\multicolumn{2}{|c|}{\bfseries Sensing Parameters} \\
\hline
$\tau_k$ & $\{30,32.55,40.71,45.49\}\text{ ns}$ \\
\hline
$\phi_k$ & $\{0,16.72,-33.72,59.09\}^{\circ}$ \\
\hline
$\theta_k$ & $\{0,30.96,-53.13,39.81\}^{\circ}$ \\
\hline
$\Re\gamma_{k,m}$ &
$\begin{gathered}
10^{-3}
\begin{bmatrix}
6.93 & 6.06 & 4.04 & 2.79\\
1.20 & 3.35 & 2.98 & 0.63\\
10.46 & 9.16 & -0.38 & 0.33\\
-6.87 & -8.76 & -0.50 & 1.87
\end{bmatrix}
\\[-1mm]
\text{Rows: }k;\text{ columns: }m
\end{gathered}$ \\
\hline
$\Im\gamma_{k,m}$ &
$\begin{gathered}
10^{-3}
\begin{bmatrix}
0.00 & 0.00 & 0.00 & 0.00\\
-8.94 & 7.15 & 4.34 & -3.58\\
1.37 & 1.13 & -6.14 & 4.24\\
-7.66 & 2.06 & 5.98 & -3.71
\end{bmatrix}
\\[-1mm]
\text{Rows: }k;\text{ columns: }m
\end{gathered}$ \\
\hline
\rowcolor{gray10pct} \multicolumn{2}{|c|}{\bfseries Algorithm Parameters} \\
\hline
$N_{\tt{MC}}$ & 1024 \\
\hline
$C_{\tt{max}}$ & 0.75 \\
\hline
$R$ & 0.5\\
\hline
\end{tabularx}
\end{table}

We primarily aim to illustrate the expected tradeoff among the three components of the channel as we incorporate multi-band. As frequency increases, we expect lower \ac{SC} power but also less \ac{DMC} spreads, implying higher decay rates (and also \ac{VMD} parameters, though not modeled in the simulation in this paper for tractability), leading to more concentrated \ac{DMC} power to the \ac{LoS} delay (and angle). For Figs. \ref{fig:crb-pt-var_alpha} to \ref{fig:simulation_PD_PFA-var_thresh}, a table of numerical parameters used is given in Table \ref{table:numerical-parameters}, where $N_{\tt{MC}}$ is the number of Monte-Carlo simulations. 

Note that the angular covariance matrices of the \ac{DMC} $\pmb{R}^{\tt{T}}_m$ and $\pmb{R}^{\tt{R}}_m$ are modeled as $\pmb{I}$ in Figs. \ref{fig:crb-pt-var_alpha} to \ref{fig:simulation_PD_PFA-var_thresh}. This is because exact modeling of these matrices would require knowledge of the system-specific frequency-dependent antenna patterns, which is more suitable in works involving experimentation \cite{bomfin2024experiment,bomfin2026icc}, and this is planned for future work. Thus, we use the typical assumption of white \ac{DMC} in space, without any specific antenna patterns, which is more tractable and sufficient for deriving the most important insights of the paper.

\subsection{System Model Analysis}

In this analysis, we provide a few intuitive examples of how the \ac{DMC}-like clutter affects the \ac{SC} estimation in multi-band \ac{ISAC}.
The goal of these results is to illustrate how the \ac{DMC} might affect different sub-bands differently, which should be taken into account in multi-band \ac{ISAC}.

\subsubsection{Channel PDP Visualization}

\begin{figure}[!t]
\centering
\includegraphics[width=3.5in]{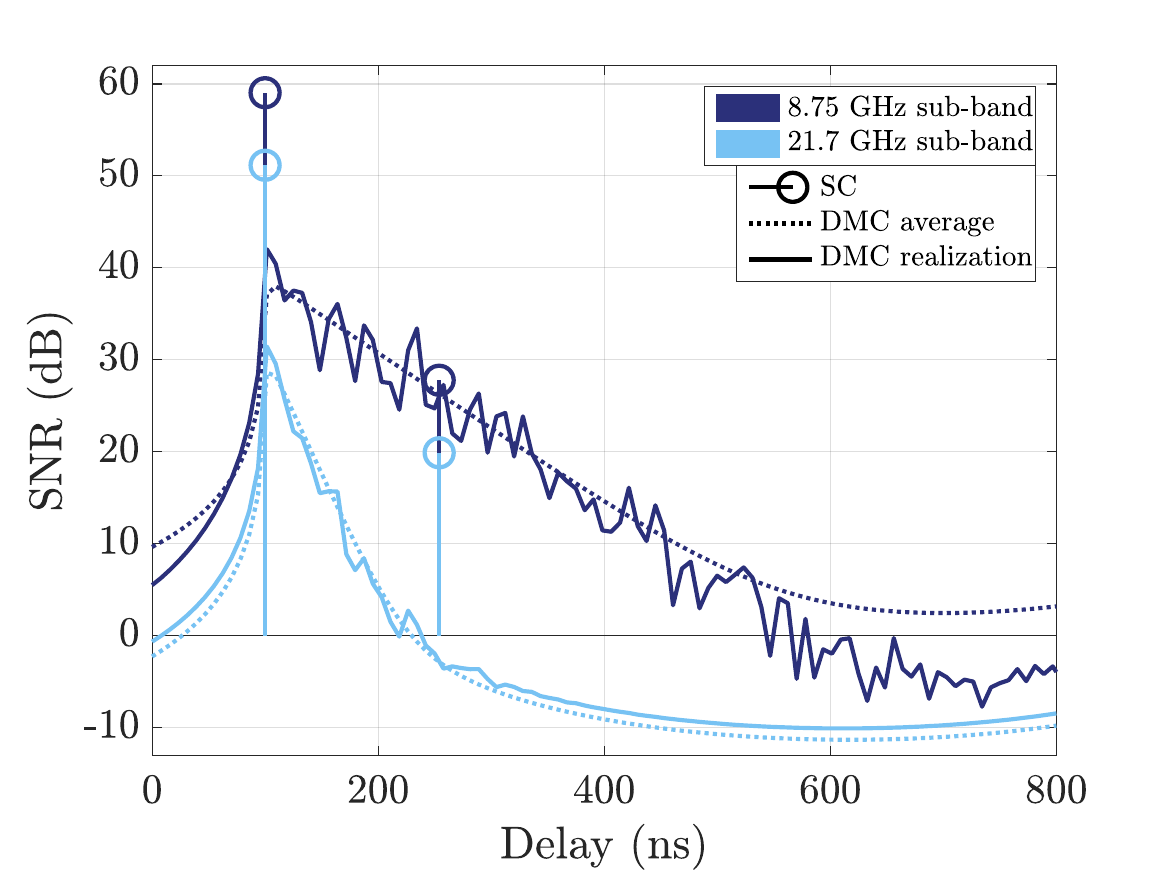}
\caption{\oscar{The \ac{PDP} of the \ac{DMC} for an 8.75 GHz sub-band with a decay rate of $\tilde{\beta} = 0.5$ and a 21.7 GHz sub-band with a decay rate of $\tilde{\beta} = 1.5$.}}
\label{fig:pdp}
\end{figure}

Fig. \ref{fig:pdp} shows the \ac{DMC} and \ac{SC} altogether in the channel \ac{PDP} when $P^{\tt{T}} = -40$ dBm/Hz and $\alpha = -20$ dB, which provides a visual representation on how the \ac{DMC} is impairing the \ac{SC} estimation. We observe that \ac{DMC} acts as multipath interference on the \ac{SC} and scales together with the \ac{SC}. As expected from the $\alpha = -20$ dB, the peak \ac{SNR} of the \ac{DMC} \ac{PDP} sits at 20 dB below the \ac{SNR} of the \ac{LoS} \ac{SC}. Thus, some \ac{SC}, such as the reflection path for the 8.75 GHz sub-band in the figure, may be shadowed under the \ac{DMC} profile. Also, due to the randomness of the \ac{DMC} profile, it may be hard to distinguish \ac{SC} from \ac{DMC} in a finite bandwidth \acp{PDP}. Finally, it is evident in this figure that although higher sub-bands suffer from lower \ac{SC} powers (e.g., for the bistatic path sitting at a delay of 254.01 ns, its \ac{SNR} on the 21.7 GHz sub-band is \oscar{7.89} dB lower than that on the 8.75 GHz sub-band), they are also blessed by the faster \ac{DMC} decay rates, which gives them lower interference.

In summary, this figure illustrates the complex interaction between \ac{SC} and \ac{DMC} profiles, where \ac{SC} with less power may be more distinguishable when \ac{DMC} interference is not dominant. Thus, this example highlights the importance of a complete analysis involving \ac{SC} and \ac{DMC} for \ac{ISAC} systems.

\subsubsection{Modified Cassini Ovals of BGG}

\begin{figure}[!t]
\centering
\includegraphics[width=3.5in]{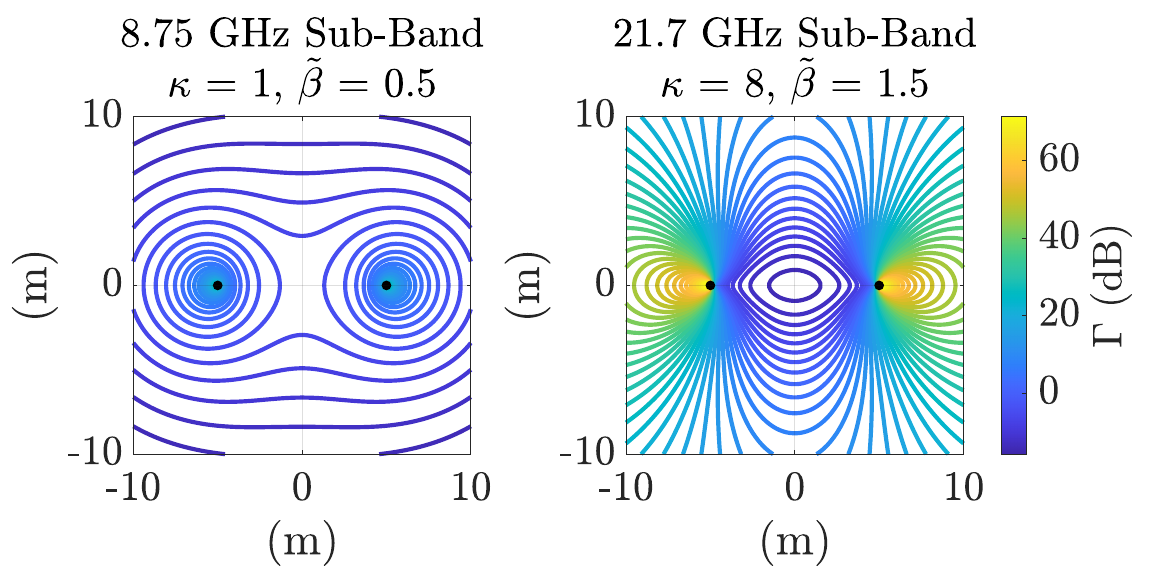}
\caption{\oscar{The modified Cassini ovals from the \ac{BGG} contours for two different sub-bands with different decay rates and \ac{VMD} parameters.}}
\label{fig:ovals_of_cassini}
\end{figure}

In the following results depicted in Fig. \ref{fig:ovals_of_cassini}, we aim at incorporating the DMC interference in a modified Cassini ovals chart, by plotting the \ac{BGG} contour when $P^{\tt{T}} = -50$ dBm/Hz and $\alpha = -5$ dB. This deviation from the conventional ovals occurs when the sub-band \ac{DMC} has a high $\kappa_m$, indicating that the angles are less spread. This observation may be counterintuitive in the context of traditional channel models, as it means that lower sub-bands, due to their \ac{DMC} \ac{PAP} being more spread out, may lose the coverage advantage typically discussed in multi-band sensing compared to higher sub-bands. This observation also suggests that in some cases (usually at higher sub-bands), the existence of more concentrated \ac{DMC} interference to the \ac{LoS} means the \ac{BGG} improves as the scatterer moves away from the \ac{Tx} or \ac{Rx}. 

In summary, these results reveal a significant feature of multi-band \ac{ISAC} systems operating in complex environments with \ac{DMC}-like clutter, where the angular \ac{DMC} profile may yield an unbalanced interference pattern. This phenomenon is particularly prominent in a multi-band due to a frequency-dependent \ac{DMC} pattern, enabling a new degree of freedom for resource optimization.

\subsection{Fundamental Limits Analysis}
\label{subsec:fundamental_limits_analysis}

In the following, we assess the \ac{CRB} derived in Section \ref{sec:fundamental-limits-and-empirical-metrics} in a single-band and multi-band setting. 
The goal of this analysis is to theoretically understand how the \ac{SC} power and \ac{DMC} clutter affect each sub-band separately, and how multi-band processing can improve the performance. \oscar{For clarity, we specifically focus on the \textit{simple scenario} in this subsection, which refers to one with two paths ($k = \{1,2\}$), two sub-bands ($f_m = \{8.75,21.7\} \text{ GHz}$), and two antennas for both the \ac{Tx} and \ac{Rx}.}

\subsubsection{CRB analysis}

\begin{figure}[!t]
\centering
\includegraphics[width=3.5in]{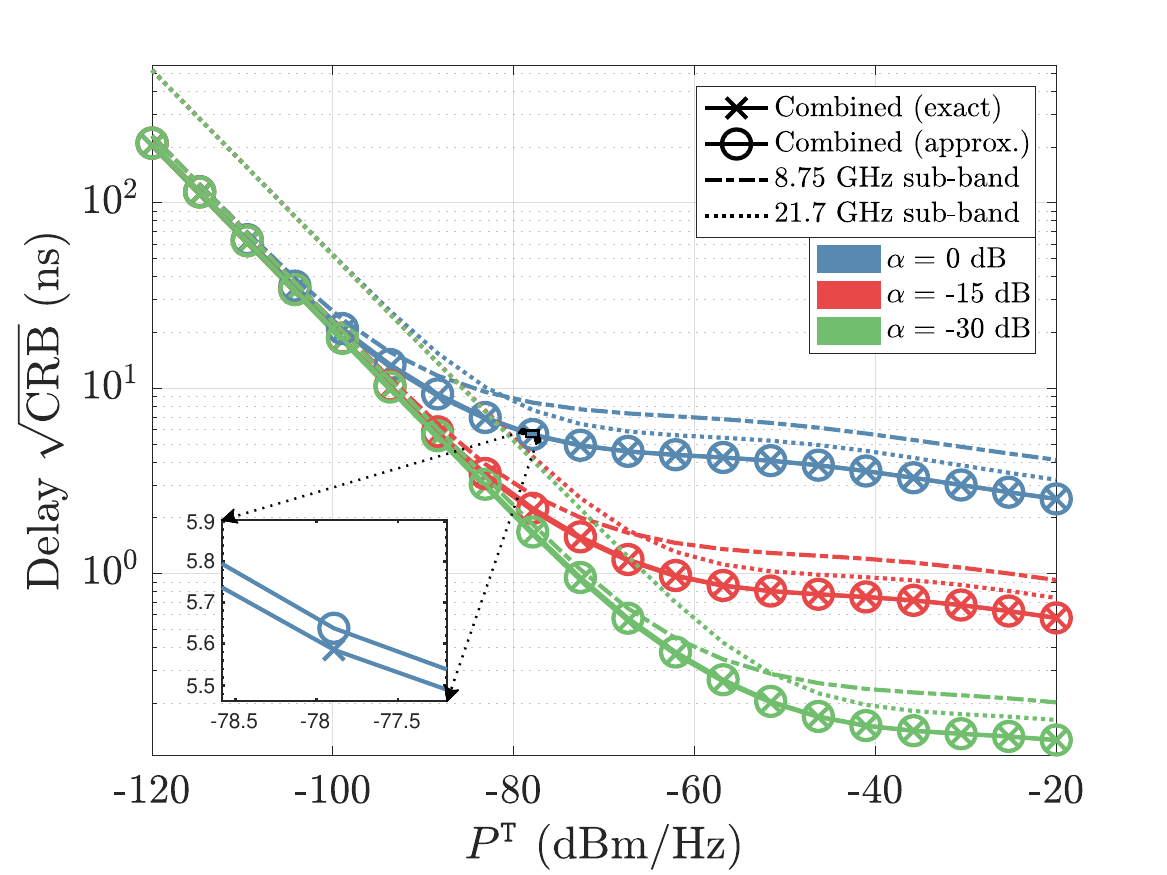}
\caption{\oscar{The delay $\sqrt{\CRB}$ vs. $P^{\tt{T}}$ for individual sub-bands and the combined multi-band under the simple scenario with two paths, two sub-bands, and a $2\times 2$ antenna configuration. Both approximate and exact \acp{CRB} are shown for the combined case.}}
\label{fig:crb-pt-var_alpha}
\end{figure}

Fig. \ref{fig:crb-pt-var_alpha} shows the \ac{CRB} of $\tau_2$, the \ac{ToA} of the scatterer in a two-path system. Firstly, we observe that the approximate \ac{CRB} of the proposed multi-band estimator closely follows the exact multi-band \ac{CRB} curve, indicating that our proposed estimator is a simple yet effective algorithm for exploiting multi-band gains. For example, for the curve with $\alpha = 0$ dB, at an exact multi-band delay $\sqrt{\CRB}$ of \oscar{5.59} ns, the difference between the approximate and the exact $\sqrt{\CRB}$ values is only \oscar{0.052} ns. Next, by inspecting individual curves, we observe that in both single-band and multi-band systems with background \ac{DMC}, the \ac{CRB} has a noise-dominated region (at lower \ac{SNR}) and a \ac{DMC}-dominated region (at higher \ac{SNR}). The noise-dominated region is defined as the region where the $\sqrt{\CRB}$ decreases linearly on a logarithmic scale. While in the \ac{DMC}-dominated region, $P^{\tt{T}}$ exhibits less and less effectiveness in decreasing the \ac{CRB}, gradually making the \ac{CRB} approach a floor. The reason behind this phenomenon is that the \ac{DMC} magnitude grows with $P^{\tt{T}}$. We also observe that lower $\alpha$ pushes the lower \ac{SNR} bound of the \ac{DMC}-dominated region higher and the \ac{CRB}-floor in the \ac{DMC}-dominated region lower, resulting in consistent drops in the logarithmic scale. For example, each 15 dB drop in $\alpha$ results in a decline in the delay $\sqrt{\CRB}$ of about \oscar{0.65} decades.

Moreover, we observe from Fig. \ref{fig:crb-pt-var_alpha} that the two sub-bands' performances ``switch" between the two \ac{SNR} regions dominated by different channel effects: in the \ac{DMC} dominated region, the 8.75 GHz sub-band has a higher \ac{CRB}, while in the noise-dominated region, the 21.7 GHz one has a higher \ac{CRB}. This behavior is because in the predominant case, we model lower path gains at higher frequencies, i.e. \oscar{$\magn{\gamma_{k,21.7}} < \magn{\gamma_{k,8.75}}$}, and at the same time, we observe more concentrated \ac{DMC} profiles at higher sub-bands, with higher decay rates and sharper angular spread. Hence, the tradeoff between \ac{SC} power and \ac{DMC} interference power is observed. While ``switching" is not guaranteed in all cases, the sole existence of such a switching phenomenon implies that one sub-band no longer dominates the estimation accuracy under all \ac{SNR} conditions, further encouraging the use of multi-band sensing. We also observe that the combined multi-band system performs better than any of the constituent sub-bands at all transmit powers. In the \ac{DMC}-dominated region, the improvements on the logarithmic scale are consistent at about \oscar{0.20} and \oscar{0.11} decades for the 8.75 and 21.7 GHz sub-bands, respectively, across varying $\alpha$ levels.

\begin{figure}[!t]
\centering
\includegraphics[width=3.5in]{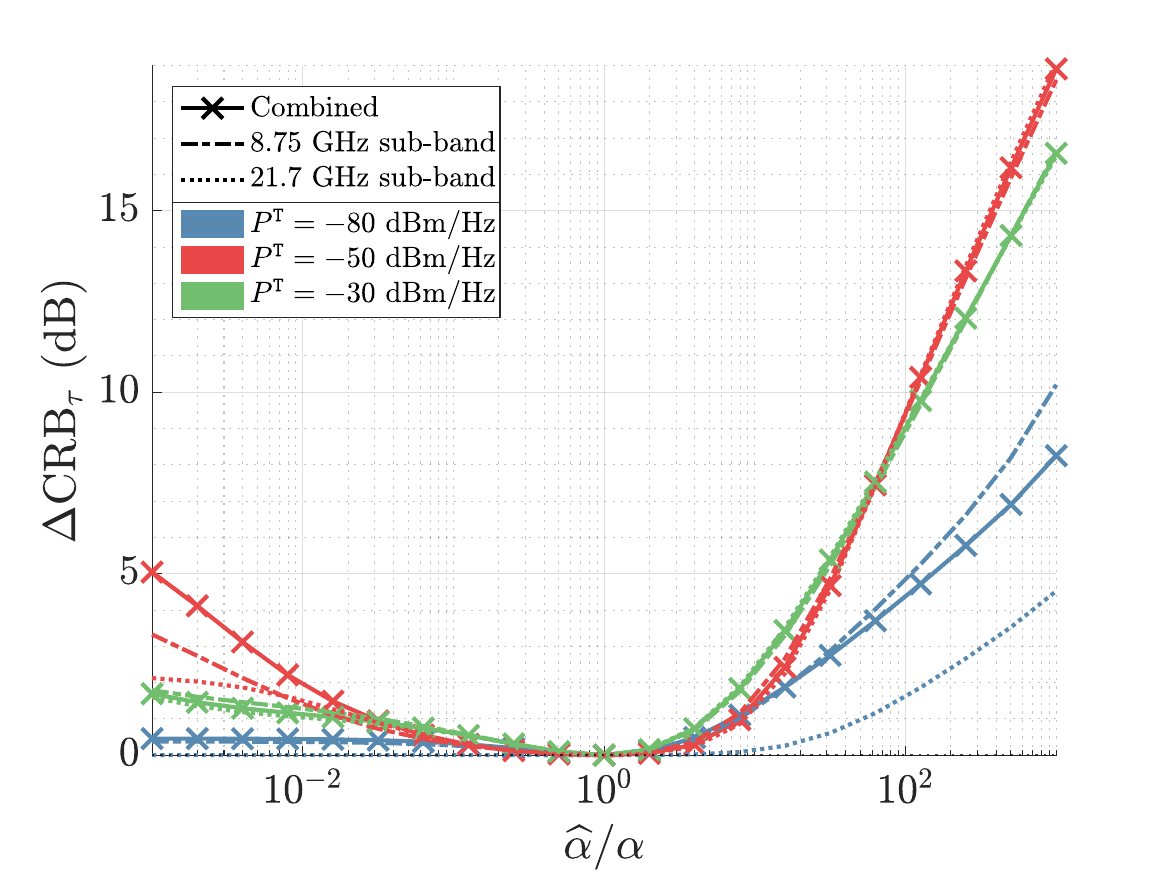}
\caption{\oscar{The $\Delta\CRB_\tau$ vs. misspecification in $\alpha$ for individual sub-bands and the combined multi-band under the simple scenario with two paths, two sub-bands, and a $2\times 2$ antenna configuration.}}
\label{fig:normalized_mcrb_ToA_vs_alpha_error}
\end{figure}

\oscar{Fig. \ref{fig:normalized_mcrb_ToA_vs_alpha_error} evaluates the sensitivity of the delay \ac{CRB} to misspecification of $\alpha$, where the covariance matrix is constructed using $\widehat{\alpha}$ instead of its true value $\alpha = -15$ dB. Note that the results show visually asymmetric trends as the degradation is plotted in the logarithmic scale. As expected, the normalized \ac{MCRB} degradation $\Delta\CRB_{\tau}$ is zero when $\widehat{\alpha}=\alpha$, and the degradation is gradual and smooth near this point. Mismatch errors of $\alpha$ within approximately $\pm5$ dB result in less than $1$ dB degradation for all considered cases. At lower $P^{\tt T}$, the effect of covariance misspecification is less pronounced because the Gaussian noise contribution is dominant. For example, for $P^{\tt T} = -80$ dBm/Hz, $\Delta\CRB_\tau \approx 0$ dB for all values with $\widehat{\alpha} < \alpha$. At higher $P^{\tt T}$, the curves corresponding to combined and single-band processing nearly overlap. For example, for $P^{\tt T} = -30$ dBm/Hz, all three $\Delta\CRB_\tau$ curves differ by less than 1 dB for all mismatch error levels considered. Thus, although their absolute \ac{CRB} values may remain different, their relative sensitivity to misspecification of $\alpha$ is approximately the same.}

\subsubsection{ESNR analysis}

\begin{figure}[!t]
\centering
\includegraphics[width=3.5in]{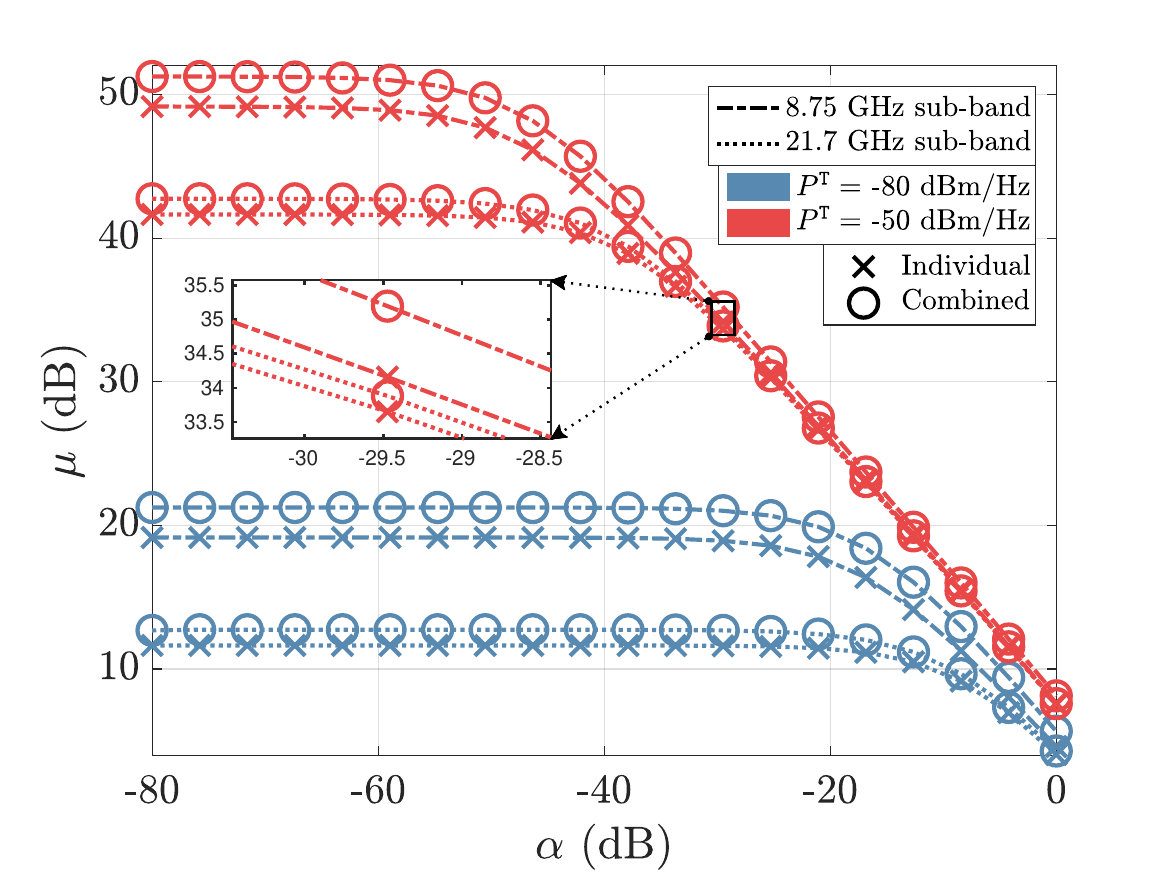}
\caption{\oscar{The \ac{ESNR} vs. $\alpha$ for individual sub-bands and the combined multi-band under the simple scenario with two paths, two sub-bands, and a $2\times 2$ antenna configuration.}}
\label{fig:esnr-alpha-var_pt}
\end{figure}

Fig. \ref{fig:esnr-alpha-var_pt}, as compared to Fig. \ref{fig:crb-pt-var_alpha}, shows a similar trend of having a \ac{DMC}-dominated region vs. a noise-dominated region but with the true \ac{ESNR} of the path associated with the scatterer at a particular sub-band in a two-path system. However, in this figure, we fix the $P^{\tt{T}}$ while varying $\alpha$. At lower $\alpha$, the floor shows the Gaussian noise dominated region, since the \ac{CRB} is not affected by changing $\alpha$ but by changing $P^{\tt{T}}$; and at higher $\alpha$, the slope shows the \ac{DMC} dominated region, since the \ac{CRB} is not affected by changing $P^{\tt{T}}$ but by changing $\alpha$. As explained before in the definition of the \ac{ESNR} metric, when \ac{ESNR} falls below a certain threshold, the \ac{CRB} value associated with that path is not meaningful, as we no longer trust that estimate. \oscar{It is hence dropped by the path-selection step of the proposed estimator.} 

We also observe from Fig. \ref{fig:esnr-alpha-var_pt} that in the combined case, the two sub-bands are always helping each other, producing higher individual estimation \acp{SNR} than separate processing. This improvement is universal across sub-bands, and is not related to which sub-band has a higher \ac{ESNR} or \ac{CRB}. For example, at $P^{\tt{T}} = -50$ dBm/Hz and $\alpha = -29.47$ dB, the \ac{ESNR} of the 8.75 GHz sub-band is improved by \oscar{1.04} dB, and that of the 21.7 GHz sub-band is enhanced by about \oscar{0.23} dB. At $P^{\tt{T}} = -50$ dBm/Hz and $\alpha = -67.37$ dB, the improvements increase to \oscar{2.09} and \oscar{1.09} dB, respectively, for the lower and higher sub-bands.

In summary, the above results revealed that performance in each sub-band depends on the \ac{Tx} power (or \ac{LoS} \ac{SNR}) region. This happens because each frequency-dependent \ac{DMC} yields different noise or \ac{DMC}-dominated regions. This brings critical insights: (i) there is no universally best sub-band across all \acp{SNR}, (ii) multi-band processing improves the overall performance. \oscar{Moreover, the \ac{MCRB} analysis shows that covariance misspecification due to $\alpha$ is less detrimental when the misspecified value is close to the ground truth.} Lastly, the \ac{CRB} analysis demonstrates that the approximate \ac{CRB} of our estimator based on a combination of single-band estimates is very close to the exact \ac{CRB}, indicating that the proposed estimator is nearly optimal.

\subsection{Simulation Results}
In this subsection, we aim at assessing the proposed multi-band estimation described in Section \ref{sec:proposed-multi-band-estimation-algorithm} against the \ac{CRB}. In particular, we show that the relatively low-complexity strategy of combining individual single-band estimates is near-optimal, corroborating the results of Fig. \ref{fig:crb-pt-var_alpha} where the approximate \ac{CRB} of the test case has practically negligible loss in relation to the exact one. \oscar{We also show the impact of system complexity and predefined algorithmic threshold values on the performance. The 8.75, 10, and 21.7 GHz sub-bands are used in the case of ``3 sub-bands", and the 15 GHz sub-band is used in addition to this set in the case of ``4 sub-bands".}

\subsubsection{RMSE Performance} 

\begin{figure}[!t]
\centering
\includegraphics[width=3.5in]{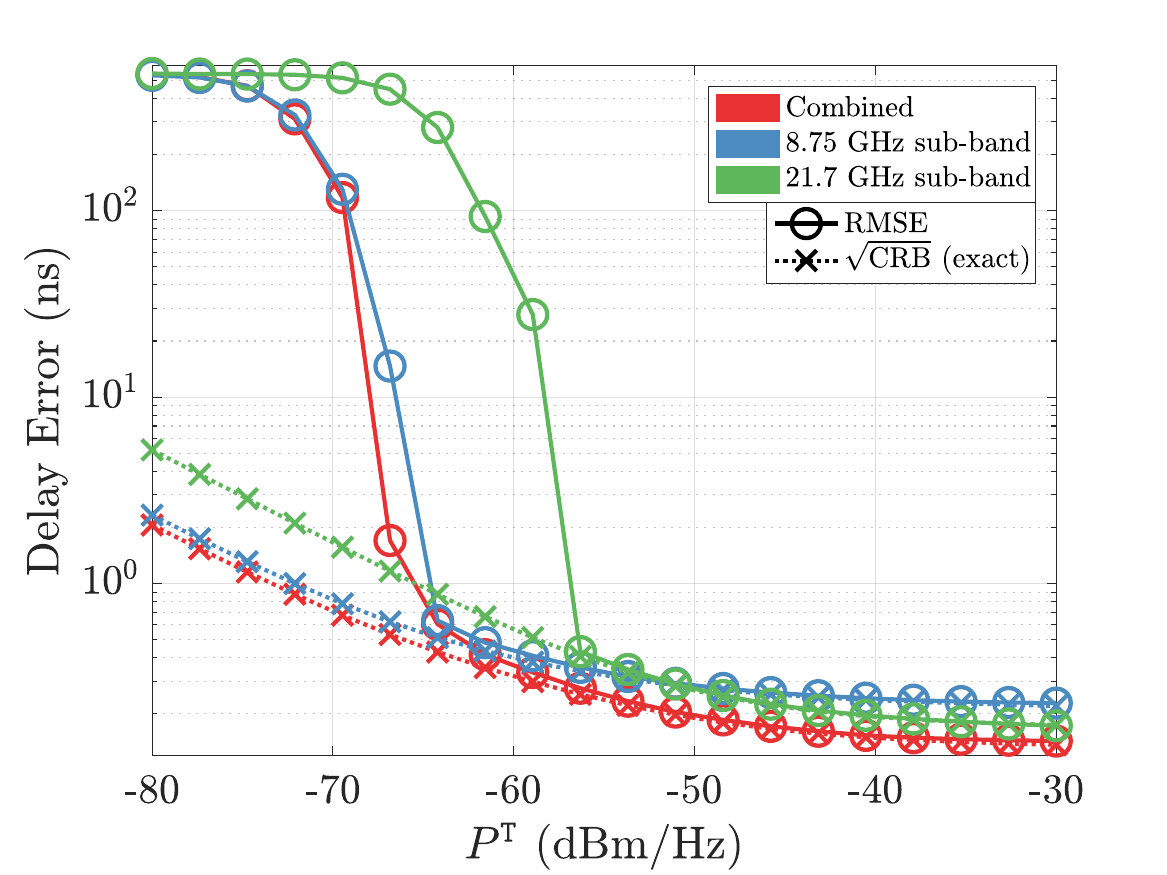}
\caption{\oscar{Numerical delay \ac{RMSE} and $\sqrt{\CRB}$ vs. $P^{\tt{T}}$ for individual sub-bands and the combined multi-band under the simple scenario with two paths, two sub-bands, and a $2\times 2$ antenna configuration.}}
\label{fig:simulation_delay_error}
\end{figure}

Fig. \ref{fig:simulation_delay_error} illustrates the estimation error in \ac{ToA} when employing the proposed estimation algorithm at $\alpha = -30$ dB, \oscar{$\varepsilon_{\tt ESNR} = 6$ dB, and $\varepsilon_{\tt prom} = 0.2$ under the \textit{simple scenario} described in subsection \ref{subsec:fundamental_limits_analysis}. Note that since we are performing \ac{JADE}, the bandwidth-derived scale $1/\BW$ does not by itself dictate the delay separability of this estimator.} First, we can see that both the multi-band estimator and the single-band estimator on the 8.75 GHz sub-band achieve the \ac{CRB} at \oscar{-61.58} dBm/Hz, while the single-band estimator on the 21.7 GHz sub-band achieves the \ac{CRB} at \oscar{-56.32} dBm/Hz. For the combined multi-band case, we compare against the exact \ac{CRB} rather than the approximate one. Note that since the simulation is run for a discrete set of $P^{\tt{T}}$ values, the exact point of achieving the \ac{CRB} cannot be directly inferred from the plot. 

Next, the gain in \ac{RMSE} performance of the combined multi-band estimator is more significant in the \ac{CRB}-achieving region as compared to the single-band ones. For example, at $P^{\tt{T}} = -30$ dBm/Hz, the delay \ac{RMSE} of the combined multi-band estimator is \oscar{0.143} ns, while those for the 8.75 GHz and 21.7 GHz single-band estimators are \oscar{0.228} ns and \oscar{0.173} ns, respectively. These improvements in delay \ac{RMSE} translate into reductions of \oscar{37.4}\% and \oscar{17.2}\% for the 8.75 GHz and 21.7 GHz single-band estimators, respectively. Furthermore, as predicted by the observations from Fig. \ref{fig:crb-pt-var_alpha}, the gain in estimation error is especially more prominent in the ``switching" region. Finally, we can see that in lower-\ac{SNR} regimes when the estimates are not reliable (not achieving the \ac{CRB}), the \ac{RMSE} curve for the combined multi-band in this regime mainly follows the sub-band with the lower \ac{RMSE}, with only minor improvements.

\begin{figure}[!t]
\centering
\includegraphics[width=3.5in]{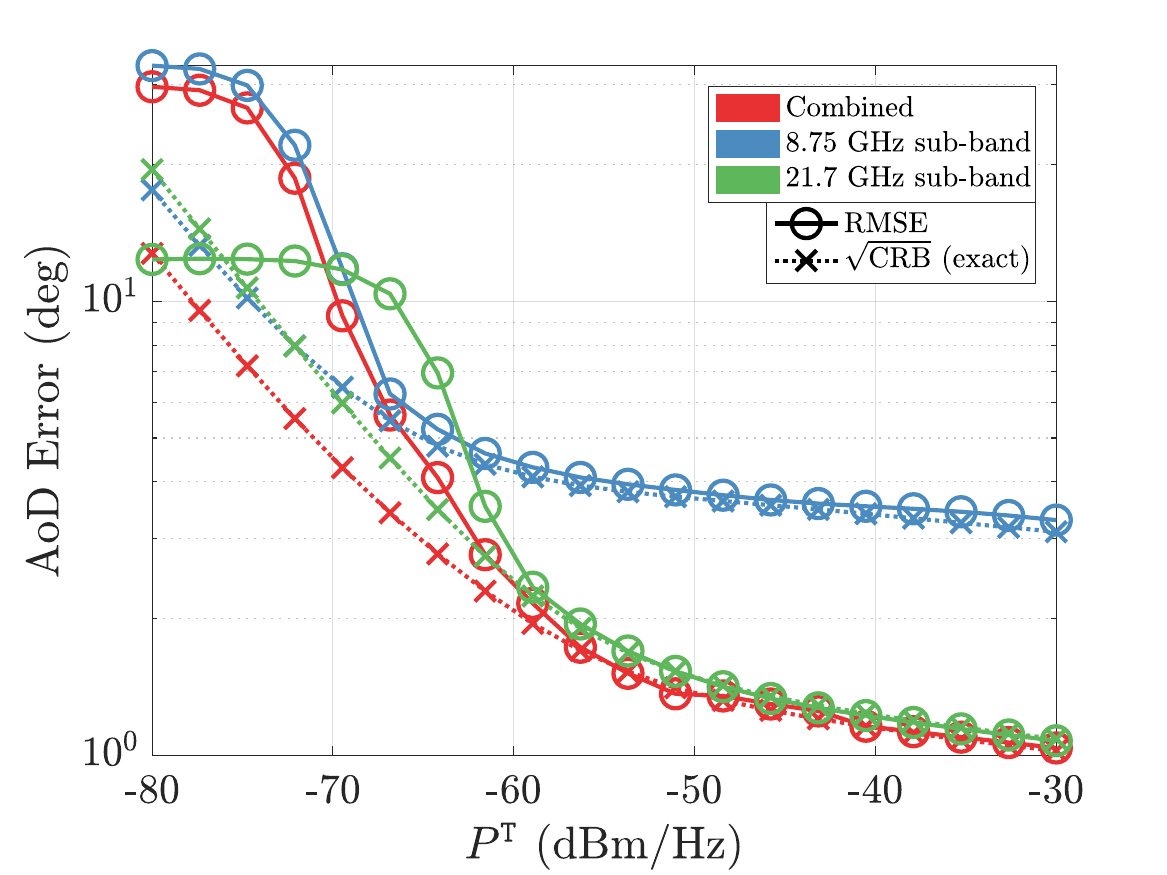}
\caption{\oscar{Numerical \ac{AoD} \ac{RMSE} and $\sqrt{\CRB}$ vs. $P^{\tt{T}}$ for individual sub-bands and the combined multi-band under the simple scenario with two paths, two sub-bands, and a $2\times 2$ antenna configuration.}}
\label{fig:simulation_AoD_error}
\end{figure}

Fig. \ref{fig:simulation_AoD_error} illustrates the estimation error in \ac{AoD} when employing the proposed estimation algorithm at $\alpha = -30$ dB, \oscar{$\varepsilon_{\tt ESNR} = 6$ dB, and $\varepsilon_{\tt prom} = 0.2$ under the \textit{simple scenario} described in subsection \ref{subsec:fundamental_limits_analysis}}. The abnormal behavior of the 21.7 GHz sub-band, where its \ac{RMSE} being capped at around $12^\circ$ for $P^{\tt{T}} = $ -80 to \oscar{-69.47} dBm/Hz and falling below the \ac{CRB} for $P^{\tt{T}} = $ -80 to \oscar{-77.37} dBm/Hz, is probably due to the existence of a false ambiguity-induced peak near the true \ac{AoD}. Apart from this, the \ac{AoD} \ac{RMSE} performance follows similar trends as compared to Fig. \ref{fig:simulation_delay_error}: In the high $P^{\tt{T}}$ region above $P^{\tt{T}} = -53.68$ dBm/Hz, all three estimators achieves the \ac{CRB}, with our combined multi-band estimator shows a gain in \ac{RMSE} \oscar{performance} compared to the constituent sub-bands; where as in the low $P^{\tt{T}}$ region, the \ac{RMSE} of the combined multi-band estimator generally follows the performance of the better constituent sub-band, with minor improvements. For example, at $P^{\tt{T}} = -40.53$ dBm/Hz, the \ac{AoD} \ac{RMSE} for the combined multi-band estimator is \oscar{$1.16^\circ$}, while that for the constituent 8.75 and 21.7 GHz sub-bands are \oscar{$3.54^\circ$} and \oscar{$1.22^\circ$}, respectively. \oscar{These improvements in \ac{AoD} \ac{RMSE} translate into reductions of 67.2\% and 5.2\% for the 8.75 GHz and 21.7 GHz single-band estimators, respectively.}

\begin{figure}[!t]
\centering
\includegraphics[width=3.5in]{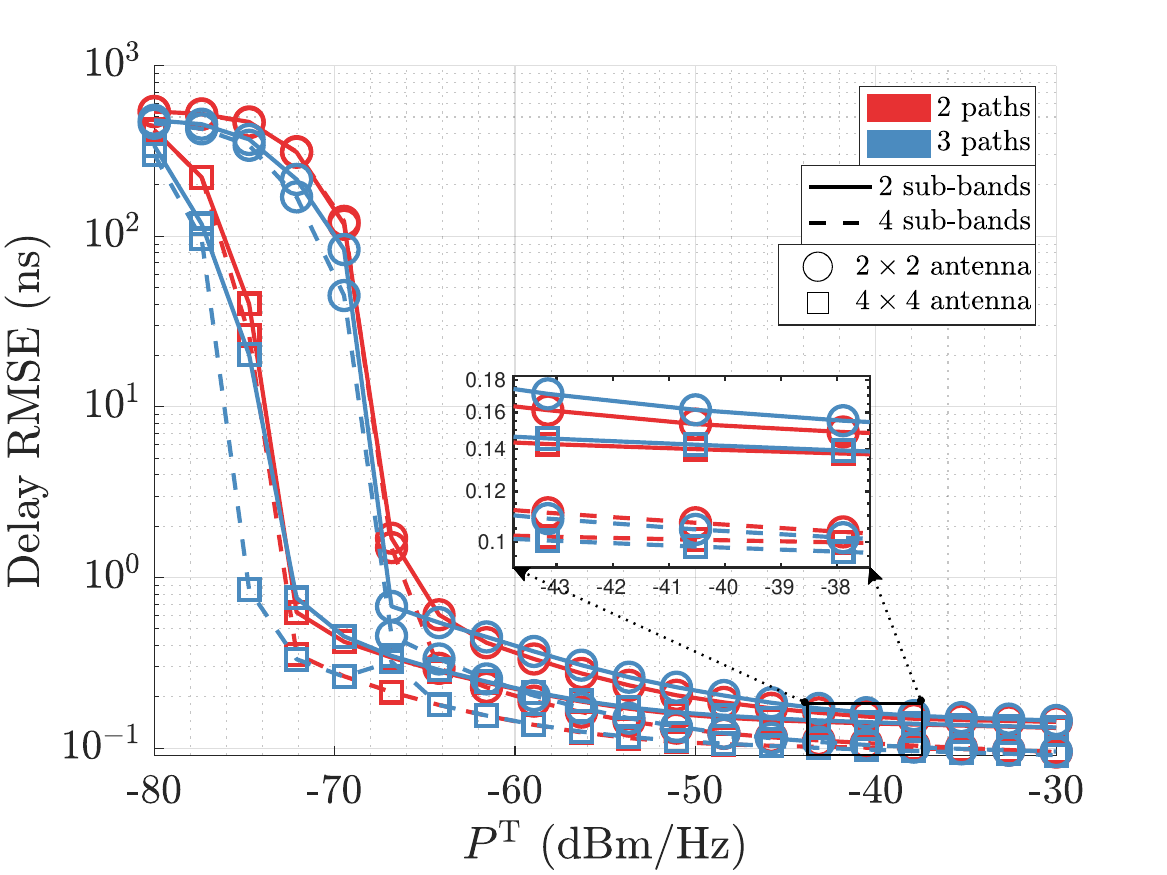}
\caption{\oscar{Numerical delay \ac{RMSE} vs. $P^{\tt{T}}$ for only the combined multi-band under more complex scenarios in terms of number of paths, sub-bands, and antennas.}}
\label{fig:simulation_RMSE-var_scenario}
\end{figure}

\oscar{Fig. \ref{fig:simulation_RMSE-var_scenario} illustrates the estimation error in the \ac{ToA} of the same $k=2$ path for only the combined multi-band case when the system complexity is increased in terms of $K$, $M$, $L^{\tt T}$, and $L^{\tt R}$ while keeping the same $\alpha = -30$ dB, $\varepsilon_{\tt ESNR} = 6$ dB, and $\varepsilon_{\tt prom} = 0.2$.  It shows that adding sub-bands mainly reduces the high-power error floor, whereas increasing the array size enables reliable acquisition at lower transmit powers and further reduces the estimation error. The $4\times4$ antenna configurations reach an \ac{RMSE} below 1 ns approximately 5.3 to 7.9 dB earlier than their $2\times2$ counterparts. For the two-sub-band cases at $P^{\tt T}=-40.53$ dBm/Hz, increasing the array size reduces the delay \ac{RMSE} from 0.153 to 0.140 ns for two paths and from 0.162 to 0.142 ns for three paths. At the same power, increasing from two to four sub-bands reduces the delay \ac{RMSE} from 0.153 to 0.107 ns for the two-path $2\times2$ case and from 0.162 to 0.105 ns for the corresponding three-path case. Increasing the number of paths has a model-dependent effect on the high-power error floor, producing only small variations among the corresponding curves in this case.} 

\begin{figure}[!t]
\centering
\includegraphics[width=3.5in]{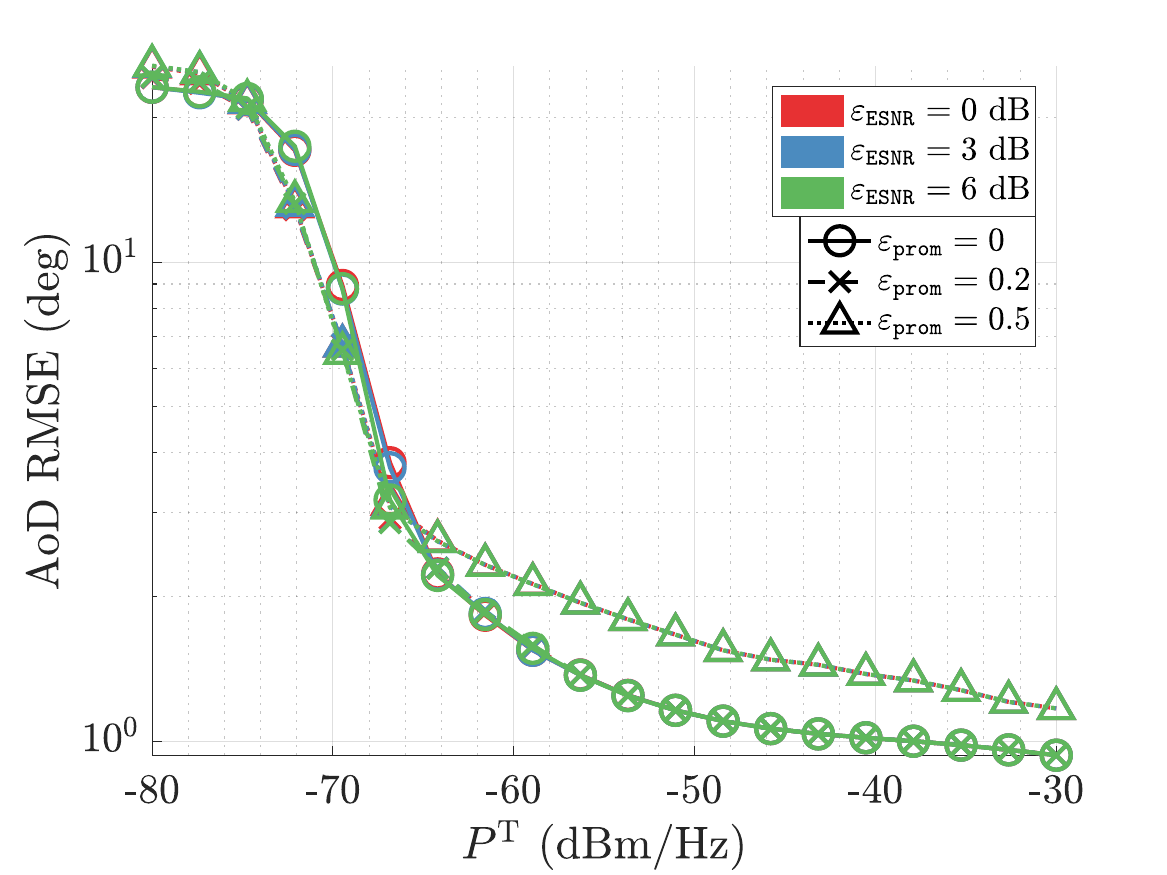}
\caption{\oscar{Numerical \ac{AoD} \ac{RMSE} vs. $P^{\tt{T}}$ at different $\varepsilon_{\tt ESNR}$ and $\varepsilon_{\tt prom}$ levels for only the combined multi-band under a scenario with three paths, three sub-bands, and a $2\times 2$ antenna configuration.}}
\label{fig:simulation_RMSE-var_thresh}
\end{figure}

\oscar{Fig. \ref{fig:simulation_RMSE-var_thresh} illustrates the estimation error in \ac{AoD} of the same $k=2$ path for only the combined multi-band case when employing the proposed estimation algorithm at $\alpha = -30$ dB under a scenario with three paths, three sub-bands, and a $2\times 2$ antenna configuration with varying $\varepsilon_{\tt ESNR}$ and $\varepsilon_{\tt prom}$. It shows that $\varepsilon_{\tt ESNR}$ barely affects the \ac{RMSE}, while $\varepsilon_{\tt prom}$ has a clearer influence on the attainable error floor: at $P^{\tt{T}}=-30$ dBm/Hz, margins of 0 and 0.2 both yield an \ac{AoD} \ac{RMSE} of approximately $0.934^\circ$, compared with $1.170^\circ$ for a margin of 0.5. Hence, an excessively strict prominence requirement discards useful cross-sub-band associations and degrades accuracy, while margins between 0 and 0.2 preserve essentially the same conditional estimation performance.}

In summary, the above results demonstrated that the proposed multi-band estimator indeed achieves the \ac{CRB}, as expected by the results of Fig. \ref{fig:crb-pt-var_alpha}. \oscar{Accuracy improves with increased array sizes or number of sub-bands, while an overly strict prominence threshold degrades conditional accuracy. The \ac{RMSE} reductions quantify accuracy improvements for already resolved paths rather than an increase in nominal delay or angular resolution.}

\subsubsection{PD-PFA Performance} 

\begin{figure}[!t]
\centering
\includegraphics[width=3.5in]{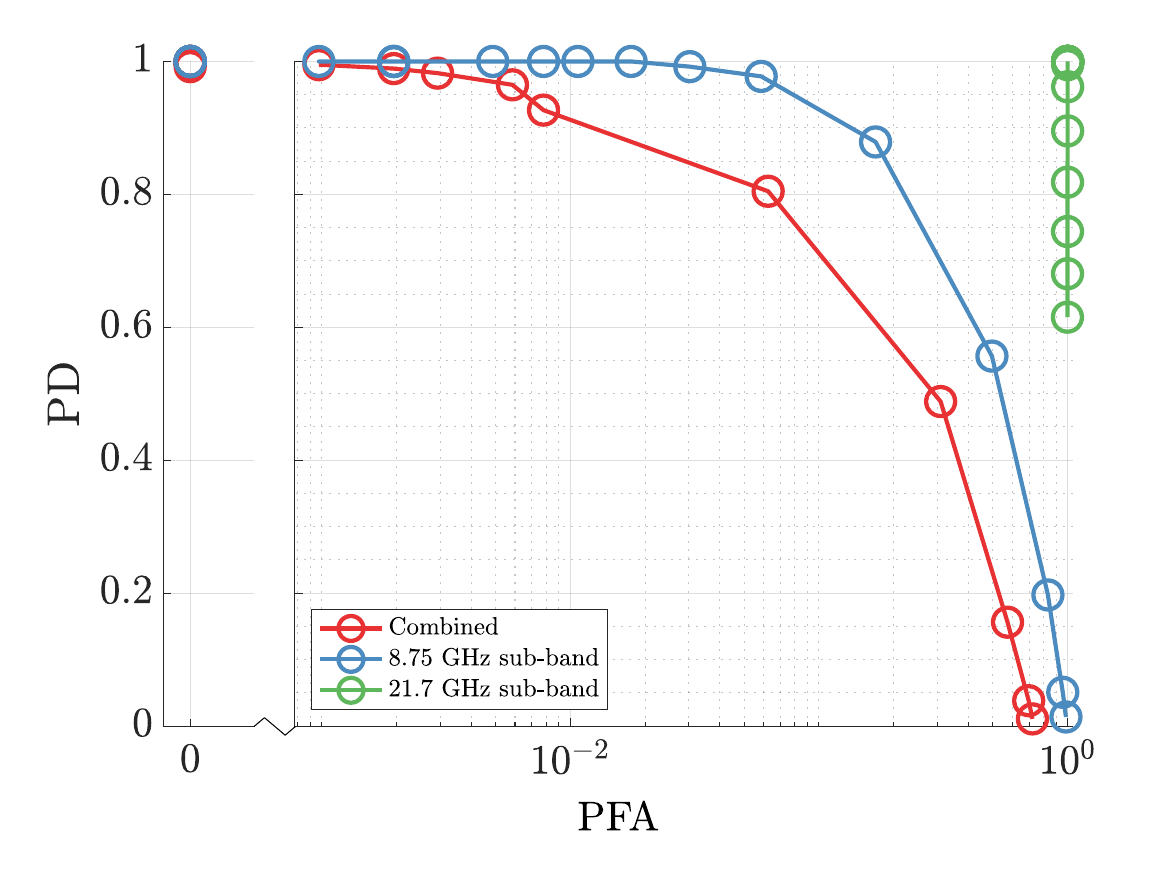}
\caption{\oscar{\Ac{PD} vs. \ac{PFA} for individual sub-bands and the combined multi-band under the simple scenario with two paths, two sub-bands, and a $2\times 2$ antenna configuration. Points with $\PFA = 0$ are separately shown.}}
\label{fig:simulation_PD_PFA}
\end{figure}

Fig. \ref{fig:simulation_PD_PFA} shows the \ac{PD}-\ac{PFA} performance curves when employing the proposed estimation algorithm at $\alpha = -30$ dB, \oscar{$\varepsilon_{\tt ESNR} = 6$ dB, and $\varepsilon_{\tt prom} = 0.2$ under the \textit{simple scenario} described in subsection \ref{subsec:fundamental_limits_analysis}}. At the same \ac{PD} level, our combined multi-band estimator has a significantly lower \ac{PFA} \oscar{compared to either single-band estimator operating on a constituent sub-band.} This is mainly because our combined multi-band estimator can resolve the grating lobes effect caused by angular ambiguity. For example, at \oscar{$\PD \approx 0.9$}, our proposed combined multi-band estimator achieves \oscar{$\PFA = 0.012$, while the 8.75 GHz and 21.7 GHz single-band estimators achieve $\PFA = 0.13$ and $\PFA = 1$, respectively. At $\PD \approx 1$, zero false alarms are observed for both the combined and 8.75 GHz estimators, whereas the 21.7 GHz estimator retains $\PFA=1$.} 

Connecting with the results shown in Fig. \ref{fig:simulation_AoD_error}, although the single-band estimator on the 21.7 GHz sub-band gives a reasonably good \ac{AoD} \ac{RMSE} in the \ac{CRB}-achieving region, that performance cannot guarantee a truthful recovery of the sensing environment due to its high \ac{PFA}. Our combined multi-band estimator, however, can resolve this issue, underscoring the need for multi-band even when \ac{RMSE} improvements are marginal.

\begin{figure}[!t]
\centering
\includegraphics[width=3.5in]{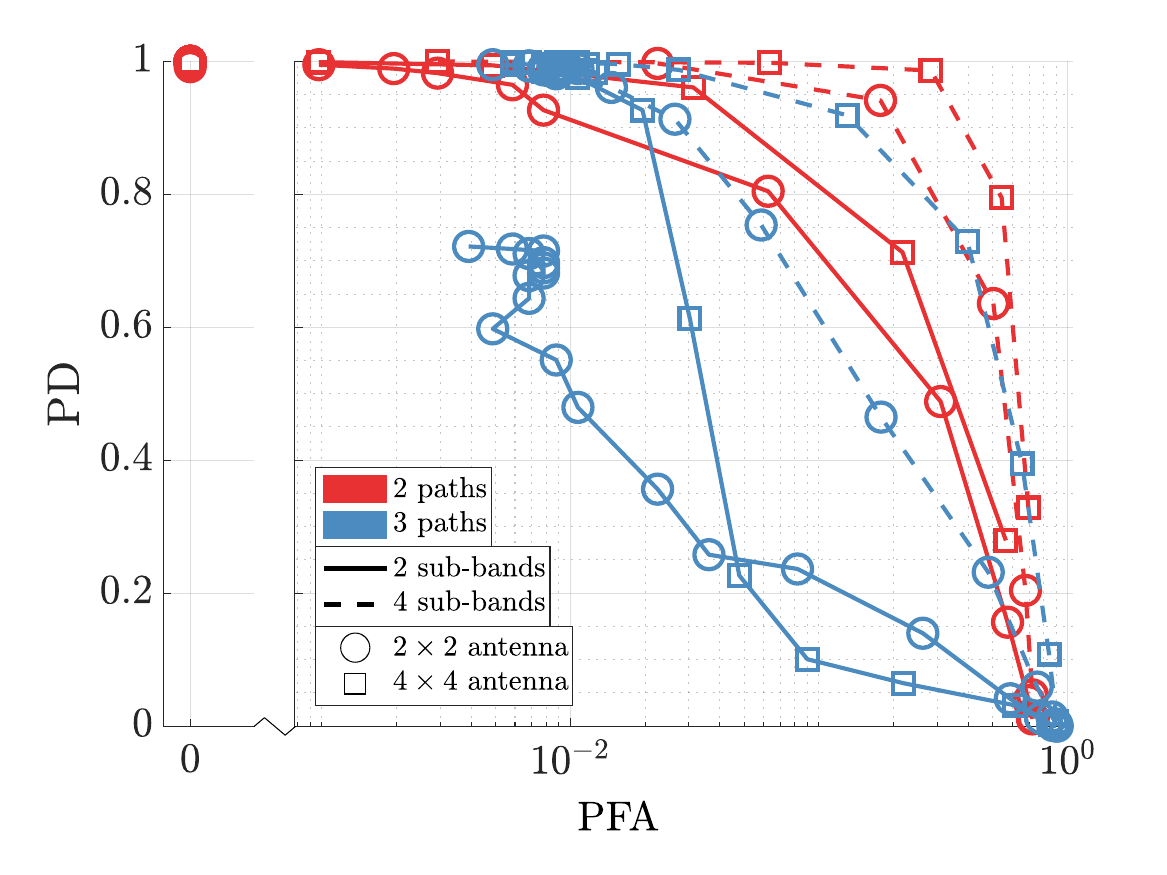}
\caption{\oscar{\Ac{PD} vs. \ac{PFA} for only the combined multi-band under more complex scenarios in terms of number of paths and sub-bands.}}
\label{fig:simulation_PD_PFA-var_scenario}
\end{figure}

\oscar{Fig. \ref{fig:simulation_PD_PFA-var_scenario} shows the \ac{PD}-\ac{PFA} performance curves for only the combined multi-band case when the system complexity is increased in terms of $K$, $M$, $L^{\tt T}$, and $L^{\tt R}$ while keeping the same $\alpha = -30$ dB, $\varepsilon_{\tt ESNR} = 6$ dB, and $\varepsilon_{\tt prom} = 0.2$. At a fixed transmit power, increasing the number of paths consistently reduces \ac{PD} and increases \ac{PFA}. For example, at $P^{\tt T}=-69.47$ dBm/Hz with two sub-bands and a $2\times2$ antenna configuration, increasing from two to three paths changes $(\PD,\PFA)$ from approximately $(0.805,0.063)$ to $(0.140,0.262)$. Increasing the number of antennas improves \ac{PD} and has a mixed, but generally beneficial, effect on \ac{PFA}: for three paths and two sub-bands at $P^{\tt T}=-64.21$ dBm/Hz, increasing the array size from $2\times2$ to $4\times4$ changes $(\PD,\PFA)$ from approximately $(0.258,0.036)$ to $(0.927,0.020)$. Increasing the number of sub-bands also improves \ac{PD}, although it tends to increase \ac{PFA} because more candidate associations are generated. For the three-path $2\times2$ case at $P^{\tt T}=-66.84$ dBm/Hz, increasing from two to four sub-bands changes $(\PD,\PFA)$ from approximately $(0.236,0.082)$ to $(0.465,0.178)$.}

\begin{figure}[!t]
\centering
\includegraphics[width=3.5in]{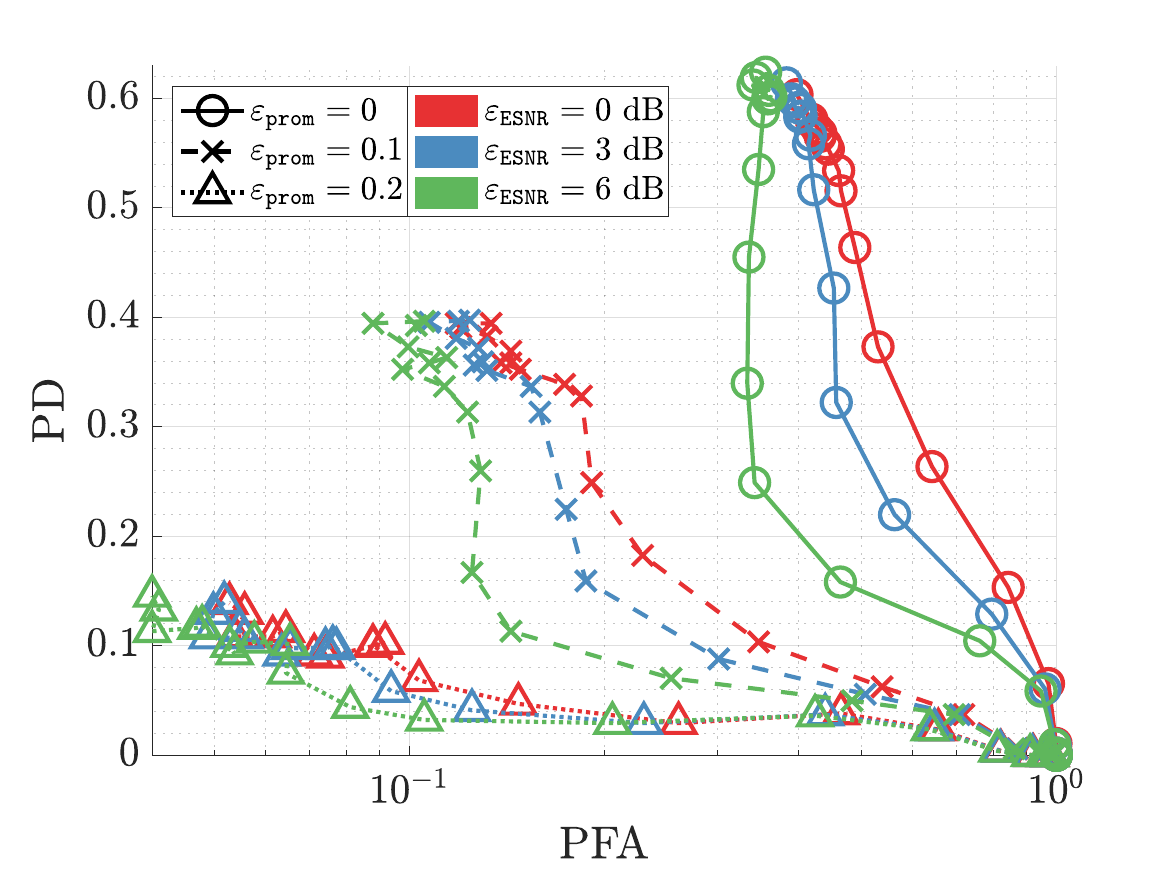}
\caption{\oscar{\Ac{PD} vs. \ac{PFA} at different $\varepsilon_{\tt ESNR}$ and $\varepsilon_{\tt prom}$ levels for only the combined multi-band under a scenario with four paths, three sub-bands, and a $2\times 2$ antenna configuration.}}
\label{fig:simulation_PD_PFA-var_thresh}
\end{figure}

\oscar{Fig. \ref{fig:simulation_PD_PFA-var_thresh} shows the \ac{PD}-\ac{PFA} performance curves for only the combined multi-band case when employing the proposed estimation algorithm at $\alpha = -30$ dB under a scenario with four paths, three sub-bands, and a $2\times 2$ antenna configuration with varying $\varepsilon_{\tt ESNR}$ and $\varepsilon_{\tt prom}$. The two thresholds control different rejection stages. Increasing $\varepsilon_{\tt ESNR}$ makes path selection stricter and generally reduces \ac{PFA} while largely preserving \ac{PD} over the tested range. At $P^{\tt{T}}=-51.05$ dBm/Hz, increasing $\varepsilon_{\tt ESNR}$ from 0 to 6 dB changes $(\PD,\PFA)$ from approximately $(0.534,0.461)$ to $(0.535,0.347)$ for $\varepsilon_{\tt prom}=0$, and from $(0.339,0.174)$ to $(0.313,0.123)$ for $\varepsilon_{\tt prom}=0.1$. In contrast, $\varepsilon_{\tt prom}$ controls the acceptance of cross-sub-band associations and produces a stronger detection--false-alarm tradeoff. At $\varepsilon_{\tt ESNR}=6$ dB and $P^{\tt{T}}=-30$ dBm/Hz, increasing $\varepsilon_{\tt prom}$ from 0 to 0.1 and 0.2 changes $(\PD,\PFA)$ from approximately $(0.624,0.355)$ to $(0.393,0.103)$ and $(0.146,0.040)$, respectively.}

In summary, the above results demonstrated that the strategy of matching across sub-bands by the Hungarian algorithm for reducing angular ambiguity due to the grating lobes effect is effective, as the proposed multi-band estimator achieves a significantly lower \ac{PFA} at the same \ac{PD} level compared to the single-band estimators on the constituent sub-bands. \oscar{Detection generally improves with more antennas or sub-bands, and more paths lead to reduced detection and increased false alarms. Over the tested range, the \ac{ESNR} threshold suppresses false alarms with little change in \ac{PD}, whereas the prominence threshold controls the stronger \ac{PD}-\ac{PFA} tradeoff.}

\section{Conclusions and Future Work}
\label{sec:conclusions-and-future-work}

\oscar{This paper developed a frequency-dependent background-\ac{DMC} model for multi-band FR3 \ac{ISAC}, analyzed its fundamental sensing limits, and proposed a scalable parameter-level fusion framework that combines single-band estimates while resolving angular ambiguities. The computational complexity analysis showed that the cost is generally dominated by parallelizable single-band estimators, while the additional fusion cost is at most quadratic in the number of sub-bands under the considered dense-solver assumptions. The \ac{PDP} analysis revealed that higher-frequency sub-bands may offset weaker \ac{SC} power through faster \ac{DMC} decay and lower interference. The modified Cassini ovals further showed that concentrated angular \ac{DMC} may result in improvements of sensing performance as the scatterer moves away from the \ac{Tx} or \ac{Rx}. The \ac{CRB} analysis identified noise- and \ac{DMC}-dominated regimes, demonstrated that no constituent sub-band is universally optimal, and showed that the proposed estimator closely approaches the exact multi-band bound. The \ac{MCRB} analysis showed less than 1 dB degradation in \ac{CRB} for errors in $\alpha$ within approximately $\pm5$ dB, and the relative sensitivity of this degradation becomes similar for both single- and multi-band at high power levels. The \ac{ESNR} analysis demonstrated how constituent sub-bands improve one another's reliability and support the rejection of unreliable estimates.}

\oscar{The simulations confirmed near-\ac{CRB} performance and reduced the delay \ac{RMSE} by 37.4\% and 17.2\% relative to the 8.75 and 21.7 GHz single-band estimators, respectively, for already-resolved paths in a representative case. Increasing the number of sub-bands primarily lowered the high-power error floor, larger arrays enabled earlier reliable acquisition, and the number of paths had a model-dependent effect on that floor, producing only small variations in the case considered. The threshold analysis showed that $\varepsilon_{\tt ESNR}$ produces only modest changes in conditional \ac{AoD} \ac{RMSE} over the tested range and can suppress false alarms, whereas $\varepsilon_{\tt prom}$ controls the stronger accuracy and detection--false-alarm tradeoffs. Finally, the \ac{PD}--\ac{PFA} analysis showed that cross-sub-band matching substantially reduces false alarms by resolving angular ambiguities; more antennas or sub-bands generally improve detection, while more paths degrade detection and increase false alarms.}

\oscar{For future work, we plan to experimentally measure the \ac{DMC} in a multi-band \ac{ISAC} scenario. We also plan to use a more physically accurate \ac{DMC} model so that parameters such as surface roughness and shape of extended targets can be estimated.}

\appendices

\section{Derivations of the Minimal Variance Linear Unbiased Estimator}
\label{appendix:derivations-of-the-minimal-variance-linear-unbiased-estimator}
For notational simplicity, in this appendix, we consider the case of some general location parameter $\psi$ estimated across sub-bands $m = 1,2,\cdots,M$. We can write the estimate per sub-band as a random variable $\hat{\psi}_m \sim \mathcal{N}\left(\psi,\sigma_m^2\right)$. We want to design the weights $\left\{w_m\right\}_{m=1}^M$ for the linear estimator $\hat{\psi} = \sum_{m=1}^M w_m \hat{\psi}_m$.

Our objective is to minimize the \ac{MSE}. But since the estimator is unbiased, the sum of weights should be 1, and the objective function is equivalently the variance of $\hat{\psi}$. In other words, we can write a problem:
\begin{equation}
\label{mvlueproblem}
\begin{aligned}
(\mathcal{P}_{\tt{MVLUE}}):
\begin{cases}
\min\limits_{w_{1:M}}
& 
\var({\hat{\psi}}) = \sum_{m=1}^M w_m^2 \sigma_m^2 \\
\textrm{s.t.} 
& \sum_{m=1}^M w_m = 1.
\end{cases}
\end{aligned}
\end{equation}
The Lagrangian is convex with
\begin{equation}
	\mathcal{L}(w_{1:M},\lambda) = \sum_{m=1}^M w_m^2 \sigma_m^2 + \lambda \left(\sum_{m=1}^M w_m - 1\right),
\end{equation}
where $\lambda$ is the Lagrangian multiplier. Hence, it is evident that we want 
\begin{equation}
	\frac{\partial \mathcal{L}}{\partial w_m} = 
	2\sigma_m^2 w_m + \lambda = 0, \forall m,
\end{equation}
which solves to $w_m \propto 1 / {\sigma_m^2}$. Substituting the constraint for obtaining an unbiased estimator, we get
\begin{equation}
	w_m = \frac{\sigma_m^{-2}}{\sum_{m=1}^M \sigma_m^{-2}}.
\end{equation}

For unbiased estimators, \ac{CRB} is equivalently the lower bound of the variances. So when the individual estimators $\hat{\psi}_m$ are efficient, we can substitute $\sigma_m^2$ with our estimated value: $\CRB_{\hat{\psi},m}$, obtaining \eqref{weights}. Since our estimated \acp{CRB} has extra randomness, we can substitute the actual $\CRB_{\psi,m}$ in place of $\sigma_m^2$ to obtain a reliable lower bound for the combined multi-band estimator, which we call \textit{approximate \ac{CRB}} as in \eqref{approxcrb}.

\bibliographystyle{IEEEtranFullAuthorDash}
\bibliography{refs}

\vspace{12pt}

\end{document}